\RequirePackage{fix-cm}
\documentclass[smallextended]{svjour3}       
\smartqed  
\usepackage{graphicx}
\usepackage{setspace}
\usepackage{rotating}
\usepackage{makecell}

\usepackage{placeins}

\usepackage{makeidx}  
\usepackage[hyphens]{url}
\usepackage{hyperref}
\usepackage{url}
\newcolumntype{?}{!{\vrule width 1pt}}
\usepackage{latexsym}
\usepackage{graphicx,paralist,amsmath,mathrsfs,multirow,amssymb,tabularx,color}
\usepackage{subfigure}
\usepackage[ruled,longend]{algorithm2e}
\usepackage{soul}
\usepackage{caption}
\usepackage[stable]{footmisc}
\newcommand*\rot{\rotatebox{90}}
\begin{document}

\title{Assessing Individual and Community Vulnerability to Fake News in Social Networks}


\author{Bhavtosh Rath \and 
		Wei Gao \and
        Jaideep Srivastava
}


\institute{B. Rath and J. Srivastava \at
              Department of Computer Science \& Engineering, University of Minnesota, USA\\
              \email{\{rathx082, srivasta\}@umn.edu}
           \and
           W. Gao \at
              School of Information Systems, Singapore Management University\\
              \email{weigao@smu.edu.sg }
}


\maketitle

\begin{abstract}
The plague of false information, popularly called \textit{fake news} has affected lives of news consumers ever since the prevalence of social media. Thus understanding the spread of false information in social networks has gained a lot of attention in the literature. While most proposed models do content analysis of the information, no much work has been done by exploring the community structures that also play an important role in determining how people get exposed to it. In this paper we base our idea on Computational Trust in social networks to propose a novel \textit{Community Health Assessment} model against fake news. Based on the concepts of neighbor, boundary and core nodes of a community, we propose novel evaluation metrics to quantify the vulnerability of nodes (individual-level) and communities (group-level) to spreading false information. 
Our model hypothesizes that if the boundary nodes trust the neighbor nodes of a community who are spreaders, the densely-connected core nodes of the community are highly likely to become spreaders. 
We test our model with communities generated using three popular community detection algorithms based on two new datasets of information spreading networks collected from Twitter. 
Our experimental results show that the proposed metrics perform clearly better on the networks spreading false information than on those spreading true ones, indicating our community health assessment model is effective.
\end{abstract}

\section{Introduction}
The use of social media platforms like Facebook, Twitter and Whatsapp is ubiquitous in modern times, making them powerful tools for information propagation and consumption. However, such goodness inevitably gets accompanied by the bad due to the innate vulnerability of human users to misinformation, which can be witnessed with the tremendous problem of \textit{fake news spreading}~\cite{kumar2018false}. \textit{Fake news} is a recently coined term that means fabricated news. \textit{It refers to newsworthy claims that may have no basis in fact, but are presented as being factually truthful.} It gets spread when someone propagates it online via various endorsements such as replying, sharing or re-posting without validating the authenticity of the content. 

There is a sheer amount of interest in the research community to understand fake news spreading, as summarized by Sharma et al.~\cite{sharma2019combating}. Our approach is orthogonal to these by focusing on {\it assessing the vulnerability of social networks to false information spreading}. 
Specifically, our focus is on people and the online communities they create, with the goal of identifying how vulnerable individuals and communities are to believing false information. We propose the Community Health Assessment model that introduces the ideas of neighbor, boundary and core nodes of a community and proposes novel metrics to quantify the vulnerability of an individual and the community itself. 
{\it From a public health perspective, determining whether a piece of news is fake or not is akin to determining whether a virus is injurious to health, while our approach is akin to determining whether an individual or community is vulnerable to being infected by the virus.} 
Thus, the proposed approach provides a complementary perspective, and can be useful in inoculating individuals and communities against spread of fake news.

To assess the vulnerability of users and their communities, we propose methods to quantify the likelihood of a boundary node of a community to believe a news item sent from its immediate neighbors, and also quantify the likelihood of a community's entire boundary node set to believing its neighborhood, i.e., the set of nodes outside the community that are connected to at least one member of the community.  Intuitively, if an external node infects a member of a community, the likelihood of the entire community to get infected increases due to high connectivity among community members. Thus, while assessing vulnerability of community, we focus on examining the influence of information propagated from external nodes into the community rather than considering the internal propagation of the news within the community.  
We evaluate our model on the propagation networks on multiple real-world information spreading networks from Twitter. 

In this paper, we make the following three novel contributions:
\begin{itemize}
\item We propose the Community Health Assessment model that initiates the ideas of neighbor, boundary and core nodes for a community structure in a social network.
\item We propose metrics that are used to quantify the vulnerability of node and community to fake news spreading from outside. Health analogy here is that fake news is akin to infection, and quantifying vulnerability is akin to assessing immunity to infection spread.
\item We present evaluation of the proposed metrics using two datasets based on a set of fact-checked news events, one from snopes.com and the other from a fact-checking website in India. 
We demonstrate that our proposed metrics can much better assess the vulnerability of social networks to fake news than regular news. To the best our knowledge, this is the first work to systematically quantify the vulnerability to online users and communities to fake news.
\end{itemize}

The rest of the paper is organized as follows: We first discuss the Related Work, followed by explanation of the Community Health Assessment model and the preliminary ideas that it builds upon. We then explain the algorithm to quantify the vulnerability metrics. Next, in the Experiments and Results section we explain the data collection process, the datasets and the metrics used for evaluation and the results. Finally, we provide concluding remarks and summarize scope of future work.

\section{Related Work}
We describe briefly prior work in three broadly related domains: \textit{Misinformation Detection}, \textit{Rumor Spreading Models} and \textit{Computational Trust}. 

\subsection{Misinformation Detection}
There has been a surge in interest among researchers over the past few years to build models to detect misinformation. Most approaches in literature model content-based and network-based characteristics of the misinformation. 
These methods include approaches to capture the style and the language of articles~\cite{horne2017just}, hyperpartisan news content~\cite{potthast2017stylometric} and cues that map language to perceived levels of credibility~\cite{mitra2017parsimonious}. Many classification models distinguishing true and false information have also been proposed. Perez-Rosas et al.~\cite{perez2017automatic} proposed a fake news detection model using linguistic features. Yang et al.~\cite{yang2012automatic} proposed a classification model using client- and location- based features extracted from micro-blogging websites. Network-based approaches that try to model the propagation structures of false information have also been proposed~\cite{jin2016news,friggeri2014rumor,rath2017retweet,ma2017detect}. The use of neural networks has gained strong attention recently. Use of convolution neural networks~\cite{borisyuk2018rosetta} and recurrent neural networks~\cite{ma2016detecting} and numerous variants of these fundamental models have shown promising results. Zhang et al.~\cite{zhang2020fakedetector} applied a graph neural network model that aggregates textual information news articles, creators and subjects to identify news veracity. Ma et al.~\cite{ma2019detect} applied generative adversarial networks to counter rumor dissemination by generating confusing training examples to challenge the discriminator of its detection capacity. Shu et al.~\cite{shu2019defend} applies attention mechanism that captures both news contents and user comments 
to propose an explainable fake news detection system. Khattar et al.~\cite{khattar2019mvae} used textual and visual information in a model variational autoencoder coupled with a binary classifier for the task of fake news detection. Lu and Li~\cite{lu2020gcan} integrated attention mechanism with graph neural networks using text information and propagation structure to identify whether the source information is fake or not.

\subsection{Rumor Spreading Models}
Infection spread models from epidemiology, namely SIR (Susceptible, Infected, Recovered)~\cite{newman2002spread}, SIS (Susceptible, Infected, Susceptible)~\cite{kimura2009efficient},  SEIZ (susceptible, exposed, infected, skeptic)~\cite{jin2013epidemiological}, SIHR (Spreaders, Ignorants, Hibernators, Removed)~\cite{zhao2013sir} and their variants~\cite{khelil2002epidemic,jin2013epidemiological,chen2020modeling,liu2016characterizing,rui2018spir} have been widely used to model information spreading, including rumors. Modelling rumor spreading as cascade structures in social networks is also well studied~\cite{friggeri2014rumor,vosoughi2018spread}. Other models have been proposed to identify the source of rumor spreading~\cite{shah2011rumors,zhu2016information}. 
Fan et al.~\cite{fan2014maximizing} proposed a model to maximize rumor containment within a fixed number of initial protectors and a given time deadline. Social networks are naturally composed of disjoint communities with relations formed within communities stronger than relations formed across communities. Focusing on such communities to understand rumor spread is a domain with a lot of research potential.

Fan et al.~\cite{fan2013least} proposed an approach to identify a minimal set of boundary nodes that would prevent spread of rumors from neighboring communities. Nguyen et al.~\cite{nguyen2012containment} proposed a community-based heuristic method to find the smallest set of highly influential nodes whose decontamination with good information would contain rumor spreading. Vosoughi et al.~\cite{vosoughi2018spread} is another closely related work that tried to empirically investigate the spread of true and false news online on a large real-world data repository from Twitter and concluded that false news spreads faster and deeper in networks compared to true news. Susceptibility of users to fake news has also been studied with a content analysis~\cite{shen2019gullible} and information diffusion~\cite{hoangvirality} perspective. What makes this work novel is that we propose content-agnostic metrics based on the underlying network structure.

\subsection{Computational Trust}
Computational social scientists have been interested to quantify the concept of trust in various domains~\cite{artz2007survey} with online social networks being one of them~\cite{sherchan2013survey}. One of the first works in the area of trust propagation in networks was by Ziegler and Lausen~\cite{ziegler2004spreading}. Some researchers have attempted to understand the role of trust in message propagation during time critical situations~\cite{imran2015processing}. Others have worked to assign scores to nodes in a trust network based on various structural aspects. Kamvar et al.~\cite{kamvar2003eigentrust} proposed \textit{Eigentrust} to rate trust scores of peers in a P2P network. Mishra and Bhattacharya~\cite{mishra2011finding} proposed an iterative matrix convergence algorithm to compute the bias and prestige of nodes in a network. Inspired by the HITS algorithm, Roy et al.~\cite{roy2016trustingness} proposed the Trust in Social Media (TSM) algorithm to compute a pair of complementary trust scores for every node in a social network, on which our vulnerability measures are built upon.


\section{Community Health Assessment Model}

\begin{figure}
\centering
\includegraphics[width=3.5in]{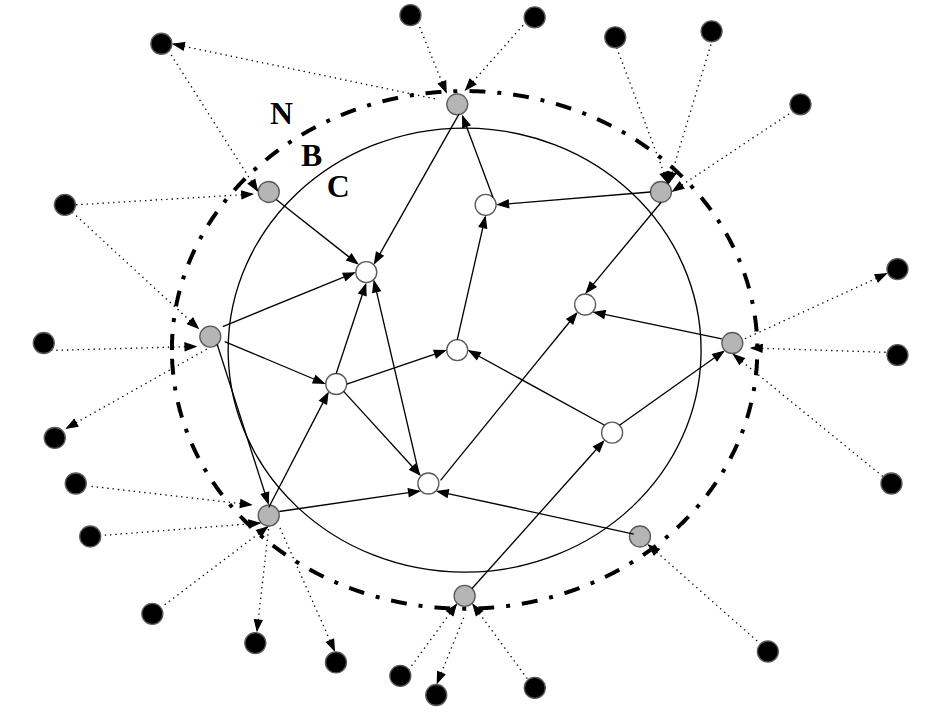}
\caption{The three levels of nodes that an piece of information can affect during its propagation. The Neighbor nodes ($\mathcal{N}$), Boundary nodes ($\mathcal{B}$) and Core nodes ($\mathcal{C}$) are represented in black, grey and white, respectively. Boundary edges (connecting neighbor and boundary nodes) are represented by dotted lines.}
\label{fig:com_health1}
\end{figure}

A social network has the characteristic property to exhibit community structures which are formed based on inter-node interactions. Communities tend to be modular groups where within-group members are highly connected, and across-group members are loosely connected. \textit{Modularity} refers to the ratio of density of edges inside a community to that of the edges outside the community. Thus, based on the edge density, members within a community would tend to have a higher degree of trust among each other than members across different communities. Also, there is variation in the level of inter-member trust across different communities due to varying modularities. If such communities are exposed to false information being propagated from neighboring nodes, the likelihood of the whole community getting infected would be high. Thus it is important to identify vulnerable communities that lie in the path of false information spreading in order to protect them and thus limit the overall influence of false information in the network. 

Motivated by this idea we propose the Community Health Assessment model. As part of the modeling, we first propose the ideas of neighbor, boundary and core nodes of a community and then propose metrics to quantify vulnerability of nodes and communities based on the fundamental measures of trust. Figure~\ref{fig:com_health1} explains the three groups of nodes with respect to a community which are affected during the process of information spreading, namely:
\begin{itemize}
    \item \textit{Neighbor nodes}: These nodes are directly connected to at least one node of the community. The set of neighbor nodes is denoted by $\mathcal{N}$. They are not a part of the community.
    \item \textit{Boundary nodes}: These are community nodes that are directly connected to at least one neighbor node. The set of boundary nodes is denoted by $\mathcal{B}$. Edges connecting neighbor nodes to boundary nodes are called boundary edges.
    \item \textit{Core nodes}: These nodes are only connected to members within the community. The set of core nodes is denoted as $\mathcal{C}$. 
\end{itemize}

\subsection{Preliminaries}

\subsubsection{Trustingness and Trustworthiness}
In social media studies, researchers have used social networks to understand how trust manifests among users. An inspiring work is the Trust in Social Media (TSM) algorithm which assigns a pair of complementary trust scores to each actor, called \textit{Trustingness} and \textit{Trustworthiness} scores~\cite{roy2016trustingness}.  \textit{Trustingness} quantifies the propensity of an actor to trust its neighbors and \textit{Trustworthiness} quantifies the willingness of the neighbors to trust the actor. The TSM algorithm takes a user network, i.e., a directed graph $\mathcal{G}(\mathcal{V},\mathcal{E})$, as input together with a specified convergence criteria or a maximum permitted number of iterations. In each iteration, for every node in the network trustingness and trustworthiness scores are computed using the equations given below:
\begin{align}
ti(v)=&\sum_{\forall x \in out(v)}\left(\frac{w(v,x)}{1+(tw(x))^s}\right) \\
tw(u)=&\sum_{\forall x \in in(u)}\left(\frac{w(x,u)}{1+(ti(x))^s}\right)
\end{align}
where $u, v, x \in \mathcal{V}$ are user nodes, $ti(v)$ and $tw(u)$ are trustingness and trustworthiness scores of $v$ and $u$, respectively, $w(v,x)$ is the weight of edge from $v$ to $x$, $out(v)$ is the set of outgoing edges of $v$, $in(u)$ is the set of incoming edges of $u$, and $s$ is the involvement score of the network. Involvement is basically the potential risk an actor takes when creating a link in the network, which is set to a constant empirically. Once the trust scores are calculated for each node in the network, TSM normalizes the scores by adhering to the normalization constraint so that both the sum of trustworthiness and the sum of trustingness of all nodes in the network equals to 1. However, a salient problem of such normalization method lies in that the scale of the scores is dependent on the size of the network. When the network is very large, the resulting scores will become extraordinarily small. To deal with the issue, min-max normalization based on the logarithm of the scores output by TSM can be used to normalize the scores into the range of (0,1]. Details about the TSM algorithm can be found in~\cite{roy2015computational}.

\subsubsection{Believability}
\textit{Believability} is an edge score derived from the Trustingness and Trustworthiness scores~\cite{rath2017retweet}. It helps us to quantify the potential or strength of directed edges to transmit information by capturing the intensity of the connection between the sender and receiver. Believability for a directed edge is computed as a function of the trustworthiness of the sender and the trustingness of the receiver.

More specifically, given users $u$ and $v$ in the context of microblogs such as Twitter, a directed edge from $u$ to $v$ exists if $u$ follows $v$. The believability quantifies the strength that $u$ trusts on $v$ when $u$ decides to follow $v$. Therefore, $u$ is very likely to believe in $v$ if:
\begin{enumerate}
\item $v$ has a high trustworthiness score, i.e., $v$ is highly likely to be trusted by other users in the network, or
\item $u$ has a high trustingness score, i.e., $u$ is highly likely to trust others. 
\end{enumerate}
So, the believability score is supposed to be proportional to the two values above, which can be jointly determined and computed as follow:
\begin{equation}\label{eq:believability}
Believability(u\rightarrow v) = tw(v) * ti(u)
\end{equation}
The idea has been previously applied in \cite{rath2017retweet} where a classification model was built to identify rumor spreaders in Twitter user network based on believability measure. 
Based on~\cite{zhao2009and}, \textit{information posted by a person the reader has deliberately selected to follow on Twitter is perceived as useful and trustworthy}, which intuitively implies that follow relation can be considered as proxy for trust.

\subsection{Vulnerability Metrics}

\begin{figure}
\centering
\includegraphics[width=\linewidth]{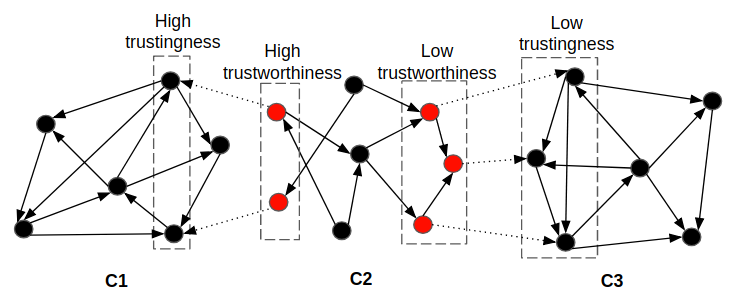}
\caption{Illustration of vulnerability to false information spreading. Red nodes are the fake news spreaders and are C1 and C3's neighbor nodes. Dotted lines denote edges connecting C1 and C3's neighbor nodes to boundary nodes.}
\label{fig:example}
\end{figure}

\textbf{Motivation:} False information generally gets very low coverage from mainstream news platforms (such as press or television), so an important factor contributing to a user's decision to spread a fake news on social media is its inherent trust on its neighbor endorsing it. On the other hand, a user would most likely to endorse a true news since it is typically endorsed by multiple credible news sources. \textit{Thus, we hypothesize that the less credible nature of false information makes it much more reliant on user's trust relationship for spreading further than true news does.} Consequently, we propose our vulnerability metrics based upon the idea of computational trust, particularly believability, for assessing the health of individuals and communities encountering false information.

\textbf{An illustrative Example:} We further illustrate the idea of our proposed vulnerability metrics through figure~\ref{fig:example}. In this example, red nodes in community C2 represent fake news spreaders, and C1 and C3 are two other communities having an identical structure. We see that C3 and C1 have 3 and 2 boundary nodes, respectively, which are directly connected to the fake news spreaders in C2 (with the edges represented through dotted lines). Based on edge count solely, one would think that C3 is more vulnerable to fake news spreading than C1. However, such speculation is not true because the boundary nodes of C3 having low trustingness scores are connected to spreaders in C2 having low trustworthiness scores, while the boundary nodes of C1, which have high trustingness scores, are connected to spreaders in C2 having high trustworthiness scores. Therefore, information are more likely to flow from C2 to C1 than to C3, i.e., C1 is more vulnerable. It is expected that our proposed metric should be able to identify C1 as more vulnerable than C3.

With the believability (Eq.~\ref{eq:believability}) which is defined on top of trustingness and trustworthiness derived from the TSM algorithm, we now derive the metrics to quantify vulnerability of nodes and communities to false information spreading. The proposed vulnerability metrics will help us quantify the likelihood of boundary nodes and communities to believe some information spreading from their neighbors. We assume that the information spreading is widespread outside of the community, i.e., at least some of the neighbor nodes of the community are spreaders. We define the node- and community-level metrics as follows:

\begin{enumerate}
\item 
\textit{Vulnerability of boundary node, $V(b)$}: This metric measures the likelihood of a boundary node $b$ to become a spreader. \textbf{ It is important to note that the method used to quantify vulnerability of a boundary node can be generalized to any node}. The metric is derived as follows: The likelihood of node $b$ to believe an immediate neighbor $n$ is a function of the trustworthiness of the neighbor $n$ ($n\in\mathcal{N}_b$, where $\mathcal{N}_b$ is the set of all neighbor nodes of $b$) and the trustingness of \textit{b}, and is quantified as $bel_{nb} = tw(n)* ti(b)$, that is, $Believability (n\rightarrow b)$. Thus, the likelihood that $b$ is \emph{not} vulnerable to $n$ can be quantified as $(1 - bel_{nb})$. Generalizing this, the likelihood of \textit{b} \emph{not} being vulnerable to all of its neighbor nodes is $\prod_{\forall n \in \mathcal{N}_b}(1 - bel_{nb})$. Therefore, the likelihood of \textit{b} to believe any of its neighbors, i.e. the vulnerability of the boundary node $b$ is computed as:
\begin{equation}
V(b) = 1 - \prod_{\forall n \in \mathcal{N}_b}(1 - bel_{nb})
\end{equation}

\item 
\textit{Vulnerability of community, $\widetilde{V}(C)$}: To compute vulnerability of community, we consider the community health perspective, i.e., vulnerability of community to information approaching from neighbor nodes (i.e., outside the community) towards the boundary node (i.e., circumference of the community). As the scenario does not include information diffusion within the community, thus the metric is independent of the core nodes of the community. This metric measures likelihood of the boundary node set of a community $C$ ($\mathcal{B}_C$) to believe an information from any of its neighbors. The metric is derived as follows: Going forward with the idea in 1), the likelihood that boundary node $b$ is \emph{not} vulnerable to its neighbors can be quantified as $(1 - V(b))$. Generalizing this to all $\textit{b}\in\mathcal{B}_C$, the likelihood that none of the boundary nodes of a community are vulnerable to their neighbors can be quantified as $\prod_{\forall b \in \mathcal{B}_C}(1 - V(b))$. Thus, the likelihood of community $C$ being vulnerable to any its neighbors, i.e., the vulnerability of the community, is defined as:
\begin{equation}
\widetilde{V}(C) = 1 - \prod_{\forall b \in\mathcal{B}_C}(1 - V(b))
\end{equation}
\end{enumerate}

The pseudo-code of algorithm to generate the vulnerability metrics is provided in Algorithm~\ref{algorithm1}.

\begin{algorithm}[t!]
  $\textbf{Input:} \quad \text{$\mathcal{G(\mathcal{V},\mathcal{E})}$: Spreader's follower-following network} $\;
  $\textbf{Output:} \quad \text{ $V(b)$: Vulnerability of each boundary node}$, and 
  $\text{$\widetilde{V}(C)$: Vulnerability of each community}$\;
  $ $\\
  $(ti, tw)_{\forall v \in \mathcal{V}} \gets \text{Trust scores using TSM($\mathcal{G}$)}$\;
  $ \phi \gets \text{Disjoint communities in $\mathcal{G}$}$\;
  $ C \gets \text{A community  s.t. $C \in\phi$}$\;
  $\mathcal{B}_C \gets \text{Set of Boundary nodes for community $C$}$\;
  $\mathcal{N}_b \gets \text{Set of Neighbor nodes for boundary node \textit{b}}$\;
   \For{each $C \in{} \phi$}
  {
     \For{each $b \in{} \mathcal{B}_C$}
    {
         \For{each $n \in{} \mathcal{N}_b$}
        {
            $bel_{nb} = tw(n) * ti(b)$
        }
        $V(b) = 1 - \prod_{\forall n \in \mathcal{N}_b}(1 - bel_{nb})$
     }
     
     $\widetilde{V}(C) = 1 - \prod_{\forall b \in{} \mathcal{B}_C}(1 - V(b))$
  }
  \caption{Vulnerability Metrics Computation
}
  \label{algorithm1}
\end{algorithm}

\section{Experiments and Results}
\subsection{Datasets and Setup}

\begin{table*}[h]
  \begin{center}
    \caption{Metadata for $DS1$.}
    \label{tab:meta_DS1}
    \resizebox{\textwidth}{!}{%
    \begin{tabular}{|c|c|c|c|l|} 
    \hline
      \textbf{Network} & \textbf{ ID} & \textbf{No. of nodes} & \textbf{No. of edges} & \textbf{Snopes link} \\
      \hline
     \multirow{4}{*}
     {\textit{Mixture}} & \textit{M1}  & 2,385,188 & 11,684,879 & \url{www.snopes.com/fact-check/nike-workers-pay-kaepernick/}  \\
     & \textit{M2}  & 3,669,213 & 7,054,734 & \url{www.snopes.com/fact-check/virginia-prisons-tampons/} \\
     & \textit{M3}  & 6,462,462 & 10,621,364 & \url{www.snopes.com/fact-check/sheriff-nike-shirt-mugshots/} \\
     & \textit{M4}   & 3,512,201 & 6,108,311 & \url{www.snopes.com/fact-check/opportunity-rovers-final-words/} \\
     \hline
     \multirow{4}{*}
     {\textit{False}} &
     \textit{F1}  & 1,883,329 & 16,658,841 & \url{www.snopes.com/fact-check/were-hate-charges-blm-kidnappers-dropped/} \\
     & \textit{F2}  & 4,981,319 & 12,625,672 & \url{www.snopes.com/fact-check/german-news-trump-nato/} \\
      & \textit{F3} & 782,209 & 12,498,122 & \url{www.snopes.com/fact-check/jussie-smollett-cnn-job/} \\
     & \textit{F4} & 503,160 & 7,797,449 & \url{www.snopes.com/fact-check/kamala-harris-jussie-smollett/} \\
    \hline
     \multirow{4}{*}
     {\textit{True}} & 
      \textit{T1}   & 10,929,291 & 14,933,611 & \url{www.snopes.com/fact-check/eva-ramon-gallegos-hpv/} \\
   & \textit{T2}   & 953,040 & 1,250,463  & \url{www.snopes.com/fact-check/nz-prime-minister-massacre-aid/} \\
     & \textit{T3}   & 2,155,927 & 3,221,985 & \url{www.snopes.com/fact-check/betsy-devos-special-olympics/} \\
     & \textit{T4}  & 1,530,958 & 2,484,553 & \url{www.snopes.com/fact-check/texas-governor-tweet-rapist/} \\\hline
    \end{tabular}
    }
  \end{center}
\end{table*}

\begin{table*}
  \begin{center}
    \caption{Metadata for $DS2$.}
    \label{tab:meta_DS2}
    \resizebox{\textwidth}{!}{%
    \begin{tabular}{|c|c|c|p{15cm}|}
    \hline
      \textbf{ ID} &  \textbf{No. of nodes} & \textbf{No. of edges} & \textbf{Link of debunked article} \\
      \hline
     \textit{N1}  & 879,854 & 2,641,513 & \url{www.altnews.in/bjp-mla-raja-singh-plagiarises-pakistan-army-song-dedicates-it-to-indian-army/} \\
     \textit{N2}  & 2,900,925 & 7,882,019 & \url{www.altnews.in/amit-malviya-targets-yogendra-yadav-via-edited-video-clip-after-tv-debate-face-off/} \\
     \textit{N3}   & 2,449,434 & 5,691,728 & \url{www.altnews.in/shivraj-singh-chouhan-tweets-clipped-video-to-portray-gaffe-by-rahul-gandhi-in-poll-speech/} \\
     \textit{N4}  & 2,663,392 & 4,082,373 & \url{www.boomlive.in/pragya-thakur-was-not-4-years-old-at-the-time-of-babri-masjid-demolition/} \\
     \textit{N5}   & 757,269 &1,880,306 & \url{www.altnews.in/2017-video-from-gujarat-shared-as-pm-narendra-modis-rally-in-hyderabad/} \\
     \textit{N6}  & 327,794 & 475,811 & \url{www.altnews.in/no-a-bjp-candidate-from-west-bengal-did-not-dress-up-as-hanuman/} \\
     \textit{N7}   & 194,075 & 304,768 & \url{www.boomlive.in/did-sp-workers-jump-the-gun-with-a-pm-akhilesh-billboard-not-quite/} \\
     \textit{N8}   & 1,600,946 & 1,731,525 & \url{navbharattimes.indiatimes.com/viral-adda/fake-news-buster/news-about-former-srilankan-cricketer-sanath-jayasuriyas-death-is-a-hoax-24858/} \\
     \textit{N9} & 720,303 & 1,215,479 & \url{smhoaxslayer.com/\%E2\%80\%8Bimported-dogs-stone-pelters-for-kasmir-or-imported-entire-video-for-inciting-communal-hatred/} \\
     \textit{N10}  & 1,197,783 & 2,036,046& \url{www.altnews.in/hindi/no-mohammad-barkat-ali-is-not-a-regular-audience-of-ndtv/} \\\hline
    \end{tabular}
    }
  \end{center}
\end{table*}

\begin{figure}[h]
\centering
\includegraphics[width=\linewidth]{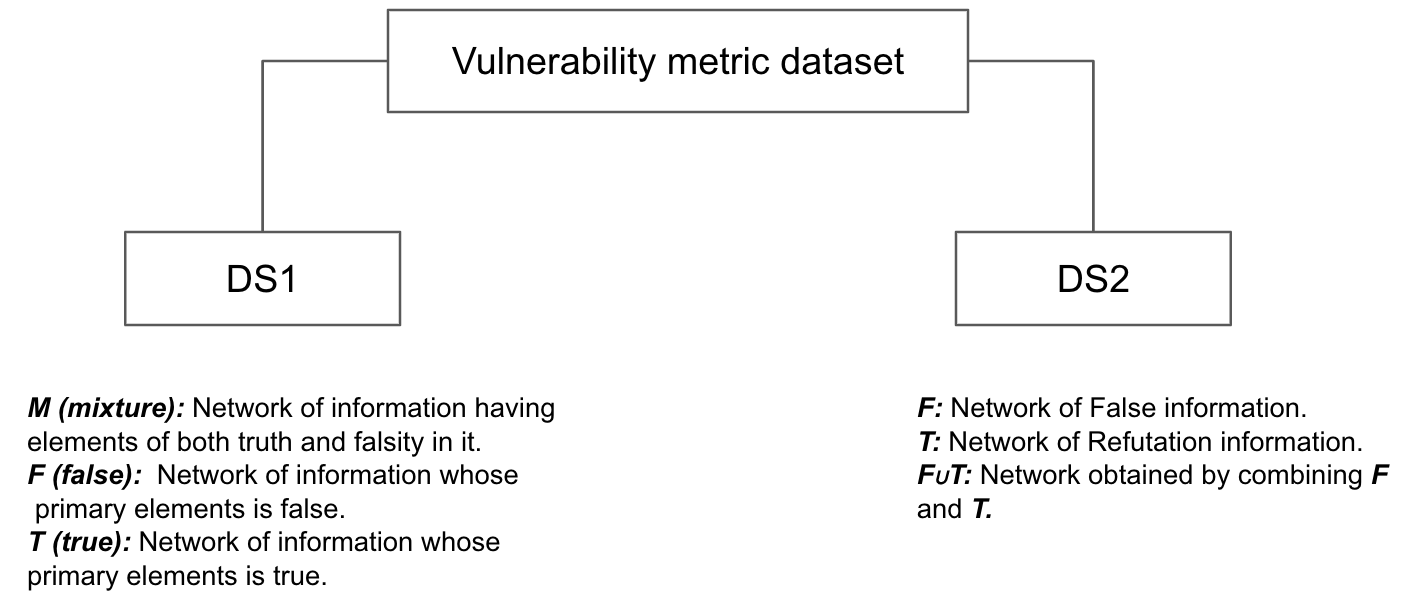}
\caption{Dataset summary.}
\label{fig:ds}
\end{figure}

We collected two sets of network datasets, namely $DS1$ and $DS2$, summarized in Figure~\ref{fig:ds} for tweets associated with news articles with confirmed ground truth of veracity from various fact checking websites. 
$DS1$ contains 12 news networks categorized into three types based on ratings of the news by snopes.com: News $M1$, $M2$, $M3$ and $M4$ are labelled as \textit{Mixture} which indicates that the news has significant elements of both truth and falsity in it, news $F1$, $F2$, $F3$ and $F4$ are labelled as \textit{False} which indicates that the primary elements of the news are basically false, and news $T1$, $T2$, $T3$ and $T4$ are labelled as \textit{True} which indicates that the primary elements of a claim are basically true. $DS2$ contains 10 news events $N1,\ldots,N10$ from fact-checking websites based in India with each news event containing a false information network denoted as $F_{N1}$,\ldots,$F_{N10}$ and its corresponding refutation information network denoted as $T_{N1}$,\ldots,$T_{N10}$. Refutation information can be defined as true information that fact checks a specific item of false information. It is created soon after a false information is debunked and tends to co-exist with the false information. We use $(F \cup T)_{N1},\ldots,(F \cup T)_{N10}$ denotes network obtained by combining false and refutation networks for specific news events.  The metadata about news events of $DS1$ and $DS2$ is described in Table~\ref{tab:meta_DS1} and Table~\ref{tab:meta_DS2}, respectively.

We identified the specific source tweet related to each information in question. For evaluation of metrics, we then identified all the spreaders of the source tweet associated with the news, which comprised of the source tweeter (identified using Twitter API) and the list of retweeters (accessible through \emph{twren.ch} or the Twitter search API). We considered the follower-following network of the spreaders obtained from Twitter API, as a proxy for social network. Code implementation and sample dataset is also provided\footnote{https://github.com/BhavtoshRath/Vulnerability-Metrics}.  

To evaluate our proposed metrics we used the collection of the twenty two different news spreading networks.   
We ran the TSM algorithm~\cite{roy2016trustingness} on follower-following network to compute the trustingness and trustworthiness scores for every node in the network. 
We then identified disjoint communities by trying three popular community detection algorithms on large networks: Louvain~\cite{blondel2008fast} solves an optimization problem that tries to maximize modularity of communities. Infomap~\cite{rosvall2008maps} algorithm is based on the principles of Information Theory. In contrast to maximizing modularity, the fundamental
approach of Infomap is to utilize flows in the graph. It uses the map equation framework, which characterizes community detection as a problem of finding a description of minimum information of a random walk process. Label Propagation~\cite{raghavan2007near} starts off by assigning a unique label to each node, and then iteratively assigns each node the label most common amongst its neighbors. As a greedy algorithm, Label Propagation is more efficient which is
linear to the number of edges in the graph. For each of the communities generated we identified the sets of boundary and neighbor nodes and then computed vulnerability metrics (see Algorithm~\ref{algorithm1}).

The network statistics based on Community Health Assessment model for $DS1$ shown in Tables \ref{tab:CD_stats_DS1}, and for false, refutation and the combined network in $DS2$ is shown in \ref{tab:CD_stats_DS2_F}, \ref{tab:CD_stats_DS2_T} and \ref{tab:CD_stats_DS2_FT}, respectively. We observe that the datasets contain varying number of communities, ranging from as low as 4/2/7 to as high as 99/2497/1637 with respect to Louvain (L)/Infomap (I)/Label Propagation (LP)\footnote{From here on, L: Louvain, I: Infomap, LP: Label Propagation in all tables.}. A general observation is that Label Propagation algorithm tends to generate more number of communities while Infomap generates fewer number of communities. Louvain gives more balanced results in terms of size and count of communities.

\begin{table*}
        \caption{Community statistics for $DS1$.}
    \label{tab:CD_stats_DS1}
        \resizebox{\textwidth}{!}{%
    \begin{tabular}{|l|c|c|c|c|c|c|c|c|c|} 
    \hline
    \multirow{2}{0.5cm}
    \textbf{\rot{Information}} & \rot{Community Detection} & \rot{\# of communities ($C$)} & \rot{Avg. \# of nodes / $C$} & \rot{Avg. \# of infected nodes / $C$} & \rot{Avg. \# of  $\mathcal{B}$ edges} & \rot{Avg. \# of $\mathcal{B}$} & \rot{Avg. \# of neighbor nodes} & \rot{Avg. \# of infected $\mathcal{B}$ nodes} & \rot{Avg. \# of infected $\mathcal{N}$ nodes} \\
    \hline
    \multirow{3}{*}{\textit{M1}}& L & 54 & 45,004 & 53 & 69,040 & 7,107 & 14,401 & 47 & 774\\
    & I & 36 & 68,148 & 81 & 5,594 & 1,778 & 1,408 & 59 & 376\\
    & IP & 786 & 3,038 & 4 & 603 & 215 & 266 & 3 & 38\\
    \hline
    \multirow{3}{*}{\textit{M2}}& L & 67 & 54,764 & 34 & 28,250 & 3,300 & 13,717 & 32 & 494\\
    & I & 5 & 733,843 & 459 & 1274 & 716 & 453 & 74 & 120\\
    & IP & 931 & 3,941 & 2 & 1,080 & 264 & 620 & 2 & 50\\
    \hline
    \multirow{3}{*}{\textit{M3}}& L & 72 & 89,756 & 39 & 20,406 & 2,878 & 11,371 & 36 & 412\\
    & I & 14 & 461,604 & 202 & 49,791 & 7,848 & 20,097 & 186 & 558\\
    & IP & 1,341 & 4,819 & 2 & 1,150 & 240 & 702 & 2 & 60\\
    \hline
    \multirow{3}{*}{\textit{M4}}& L & 99 & 35,477 & 27 & 10,606 & 2,285 & 2,996 & 23 & 484\\
    & I & 37 & 94,924  & 72 & 16,081 & 3,933 & 3,764 & 66 & 480\\
    & IP & 1,637 & 2,146 & 2 & 709 & 191 & 292 & 2 & 50\\
    \Xhline{3\arrayrulewidth}
    \multirow{3}{*}{\textit{F1}}& L & 28 & 67,262 & 103 & 218,939 & 14,547 & 34,442 & 99 & 1,028\\
    & I & 8 & 235,416 & 360 & 1,482 & 775 & 616  & 81 & 143\\
    & IP & 480 & 3,924 & 6 & 933 & 340 & 455 & 5 & 40\\
    \hline
    \multirow{3}{*}{\textit{F2}}& L & 50 & 99,626 & 57 & 51,664 & 5,793 & 21,101 & 51 & 660\\
    & I & 4 & 1,245,330 & 708 & 1,454 & 760 & 637 & 89  & 118\\
    & IP & 677 & 7,358 & 4 & 2,318 & 396 & 1,542 & 4 & 84\\
    \hline
    \multirow{3}{*}{\textit{F3}}& L & 15 & 52,147 & 31 & 417,933 &  16,382 & 52,259 & 31 & 365\\
    & I & 133 & 5,881 & 3 & 6,722 & 1,075 & 3,225 & 3 & 157\\
    & IP & 15 & 52,147 & 31 & 5,227 & 2,285 & 2,514 & 24 & 83\\
    \hline
    \multirow{3}{*}{\textit{F4}}& L & 15 & 33,544 & 19 & 338,248 &  13,848 & 56,711 & 19 & 246\\
    & I & 38 & 13,241 & 8 & 11,255 & 2,182 & 5,484 & 8 & 171\\
    & IP & 7 & 71,880 & 41 & 1,779 & 992 & 744 & 22 & 64\\
    \Xhline{3\arrayrulewidth}
    
    \multirow{3}{*}{\textit{T1}}& L & 47 & 232,538 & 59 & 47,189 & 2,171  & 42,783 & 39 & 246\\
    & I & 34 & 321,450 & 82 & 5,792 & 1,390 & 2,261 & 52 & 189\\
    & IP & 1,283 & 8,519 & 2 & 2,151 & 202 & 1,724 & 2 & 54\\
    \hline
    \multirow{3}{*}{\textit{T2}}& L & 37 & 25,758 & 5 & 4,150 & 509 & 3,095 & 3 & 36\\
    & I & 9 & 105,893 & 22 & 5,650 & 1,418 & 1,777 & 17 & 60\\
    & IP & 159 & 5,994 & 1 & 1,102 & 189 & 752 & 1 & 25\\
    \hline
    \multirow{3}{*}{\textit{T3}}& L & 27 & 79,849 & 26 & 10,135  & 1,942 & 5,251 & 18 & 180\\
    & I & 629 & 3,428 & 1 & 1,266 & 161 & 641 & 1 & 124 \\
    & IP & 209 & 10,315 & 3 & 1,138 & 303 & 584 & 3 & 46\\
    \hline
    \multirow{3}{*}{\textit{T4}}& L & 89 & 17,202 & 12 & 4,511 & 908 &  1,502 & 10 & 205\\
    & I & 1,206 & 1269 & 1 & 544 & 92 & 271 & 1 & 99\\
    & IP & 797 & 1,921 & 1 & 723 & 164 & 279 & 1 & 53\\
    \Xhline{\arrayrulewidth}
   \end{tabular}
    }
\end{table*}

\begin{table*}
        \caption{Community statistics for false information in $DS2$.}
    \label{tab:CD_stats_DS2_F}
        \resizebox{\textwidth}{!}{%
    \begin{tabular}{|l|c|c|c|c|c|c|c|c|c|} 
    \hline
    \multirow{2}{0.5cm}
       \textbf{\rot{Information}} & \rot{Community Detection} & \rot{\# of communities ($C$)} & \rot{Avg. \# of nodes / $C$} & \rot{Avg. \# of infected nodes / $C$} & \rot{Avg. \# of  $\mathcal{B}$ edges} & \rot{Avg. \# of $\mathcal{B}$} & \rot{Avg. \# of neighbor nodes} & \rot{Avg. \# of infected $\mathcal{B}$ nodes} & \rot{Avg. \# of infected $\mathcal{N}$ nodes} \\
    \hline
    \multirow{3}{*}{\textit{$F_{N1}$}} & L & 37 & 23,935 & 25 & 9,654 &  1,482 & 3,116 & 20 & 163\\
    & I & 3 & 295,199 & 314 & 17,466 & 4,786 & 3,224 & 159 & 373\\
    & IP & 220 & 4,025 & 4 & 1,299 & 322 & 623 & 4 & 46  \\
    \hline
    \multirow{3}{*}{\textit{$F_{N2}$}}& L & 66 & 39,510 & 69 & 35,877 &  2,274 & 16,655 & 62 & 562\\
    & I & 6 & 434,605 & 759 & 1 & 1 & 1 & 1 & 1\\
    & IP & 280 & 9,313 & 16 & 2,148 & 250 & 1,571 & 9 &  40 \\
    \hline
    \multirow{3}{*}{\textit{$F_{N3}$}}& L & 53 & 44,215 & 65 & 23,464 &  2,774 & 8,280 & 59 & 443\\
    & I & 2497 & 956 & 1 & 926 & 67 & 558 & 1 & 117\\
    & IP & 313 & 7,628 & 11 & 1,519 & 347 & 955 & 7 & 40\\
    \hline
    \multirow{3}{*}{\textit{$F_{N4}$}}& L & 37 & 55,031 & 24 & 8,758 &  1,102 & 6,539 & 20 & 144\\
    & I & 2 & 1,018,081 & 447 & 3,744 & 1,123 & 1,959 & 75 & 84\\
    & IP & 214 & 9,515 & 4 & 1,918 & 299 & 1,356 & 4 & 43\\
    \hline
    \multirow{3}{*}{\textit{$F_{N5}$}}& L & 47 & 11,037 & 17 & 10,685 &  1,617 & 3,319 & 17 & 234\\
    & I & 738 & 703 & 1 & 1,107 & 107 & 587 & 1 & 155 \\
    & IP & 119 & 4,359 & 7 & 1,646 & 426 & 848 & 5 & 49\\
    \hline
    \multirow{3}{*}{\textit{$F_{N6}$}}& L & 26 & 10,653 & 6 & 2,140 &  422 & 1,085 & 5 & 50\\
    & I & 4 & 69,246 & 40 & 1,734 & 584 & 564 & 17 & 64\\
    & IP & 97 & 2,856 & 2 & 768 & 152 & 379 & 2 & 29 \\
    \hline
    \multirow{3}{*}{\textit{$F_{N7}$}}& L & 20 & 7,230 & 4 & 1,261 & 381 & 324 & 3 & 27\\
    & I & 117 & 1,236 & 1 & 337 & 74 & 92 & 1 & 29  \\
    & IP & 35 & 4,131 & 2 & 724 & 245 & 207 & 2 & 16 \\
    \hline
    \multirow{3}{*}{\textit{$F_{N8}$}}& L & 17 & 23,188 & 7 & 1,479 &  308 & 911 & 5 & 49\\
    & I & 4 & 98,551 & 30 & 3,439 & 1,060 & 1,253 &  17 & 60\\
    & IP & 83 & 4,749 & 1 & 494 & 117 & 200 & 1 & 22\\
    \hline
    \multirow{3}{*}{\textit{$F_{N9}$}}& L & 43 & 11,092 & 11 & 3,673 &  802 & 1,219 & 10 & 135\\
    & I & 487 & 979 & 1 & 538 & 77 & 224 & 1 & 90\\
    & IP & 162 & 2,944 & 3 & 830 & 221 & 356 & 3 & 32 \\
    \hline
    \multirow{3}{*}{\textit{$F_{N10}$}}& L & 55 & 19,506 & 22 & 5,681 &  853 & 2,455 & 20 & 200\\
    & I & 1,045 & 1,027 & 1 & 570 & 52 & 281 & 1 & 98 \\
    & IP & 216 & 4,967 & 6 & 1,066 & 220 & 641 & 5 & 33\\
       \Xhline{\arrayrulewidth}
   \end{tabular}
    }
\end{table*}

\begin{table}
        \caption{Community statistics for refutation information in $DS2$.}
    \label{tab:CD_stats_DS2_T}
        \resizebox{\textwidth}{!}{%
    \begin{tabular}{|l|l|c|c|c|c|c|c|c|c|} 
    \hline
    \textbf{\rot{Information}} & \rot{Community Detection} & \rot{No. of communities ($C$)} & \rot{Avg. No. of nodes $/$ $C$} & \rot{Avg. No. of infected nodes $/$ $C$} & \rot{Avg. No. of  $\mathcal{B}$ edges} & \rot{Avg. No. of $\mathcal{B}$} & \rot{Avg. No. of neighbor nodes} & \rot{Avg. No. of infected $\mathcal{B}$ nodes} & \rot{Avg. No. of infected $\mathcal{N}$ nodes} \\
    \hline
    \multirow{3}{*} {\textit{$T_{N1}$}} & L & 40 & 11,338 & 10 & 5,856 &  856 & 3,018 & 8 & 96\\
    & I & 2 & 226,769 & 200 & 1,260 & 606 & 151 & 37 & 58  \\
    & IP & 154 & 2,945 & 3 & 1,564 & 274 & 1,019 & 2 & 39\\
    \hline
\multirow{3}{*}{\textit{$T_{N2}$}}& L & 47 & 9,226 & 10 & 3,562 & 540 & 1,648 & 9 & 103\\
    & I & 472 & 919 & 1 & 581 & 58 & 327 & 1 & 80\\
    & IP & 167 & 2,597 & 3 & 1,042 & 169 & 641 & 3 & 34\\
    \hline
 \multirow{3}{*}{\textit{$T_{N3}$}}& L & 15 & 86,491 & 32 & 7,305 & 987  & 5,160 & 10 & 67\\
    & I & 457 & 2,839 & 1 & 757 & 64 & 437 & 1 & 80 \\
    & IP & 84 & 15,445 & 6 & 1,497 & 260 & 1,032 & 5 & 29\\
    \hline
\multirow{3}{*}{\textit{$T_{N4}$}}& L & 45 & 23,522 & 11 & 5,399 &  590 & 3,950 & 10 & 102\\
    & I & 523 & 2,024 & 1 & 740 & 58 & 502 & 1 &77 \\
    & IP & 214 & 4,946 & 2 & 1,211 & 167 & 827 & 2 & 39\\
    \hline
\multirow{3}{*}{\textit{$T_{N5}$}}& L & 15 & 17,513 & 6 & 4,895 & 305 & 4,376 & 2 & 28\\
    & I & 2 & 131,346 & 46 & 5,650 & 936 & 2,874 &  5 & 45\\
    & IP & 40 & 6,567 & 2 & 1,769 & 159 & 1,449 & 2 &  19 \\
    \hline
\multirow{3}{*}{\textit{$T_{N6}$}}& L & 9 & 7,458 & 4 & 772 & 220 & 333 & 3 & 10\\
    & I & 103 & 652 & 1 & 106 & 25 & 48 & 1 & 16 \\
    & IP & 26 & 2,582 & 1 & 376 & 112 & 99 & 1 & 13 \\
    \hline
\multirow{3}{*}{\textit{$T_{N7}$}}& L & 20 & 3,067 & 6 & 1,280 & 267 & 478 & 5 & 38\\
    & I & 2 & 30,666 & 57 & 1,636 & 871 & 370 & 26 & 22\\
    & IP & 49 & 1,252 & 2 & 648 & 149 & 290 & 2 & 21 \\
    \hline
\multirow{3}{*}{\textit{$T_{N8}$}}& L & 4 & 310,826 & 23 & 2,152 &  233 & 1,723 & 7 & 31\\
    & I & 2 & 621,653 & 47 & 1,968 & 601 & 943 & 15 & 42\\
    & IP & 64 & 19,427 & 1 & 465 & 98 & 208 & 1 & 21\\
    \hline
\multirow{3}{*}{\textit{$T_{N9}$}}& L & 20 & 13,821 & 5 & 1,482 & 324 & 836 & 4 & 37\\
    & I & 3 & 92,143 & 32 & 233 & 81 & 132 & 9 & 16\\
    & IP & 78 & 3,544 & 1 & 564 & 120 & 244 & 1 & 28\\
    \hline
\multirow{3}{*}{\textit{$T_{N10}$}}& L & 5 & 29,757 & 7 & 1,098 & 283 & 643 & 5 & 20\\
    & I & 49 & 3,036 & 1 & 214 & 54 & 84 & 1 & 14\\
    & IP & 31 & 4,800 & 1 & 347 & 119 & 90 & 1 & 13 \\
    \hline
   \end{tabular}
    }
\end{table}

\begin{table*}
    \caption{Community statistics for false and refutation information network combined in $DS2$.}
    \label{tab:CD_stats_DS2_FT}
        \resizebox{\textwidth}{!}{%
    \begin{tabular}{|l|c|c|c|c|c|c|c|c|c|} 
    \hline
    \multirow{2}{0.5cm}
    \textbf{\rot{Information}} & \rot{Community Detection} & \rot{\# of communities ($C$)} & \rot{Avg. \# of nodes / $C$} & \rot{Avg. \# of infected nodes / $C$} & \rot{Avg. \# of  $\mathcal{B}$ edges} & \rot{Avg. \# of $\mathcal{B}$} & \rot{Avg. \# of neighbor nodes} & \rot{Avg. \# of infected $\mathcal{B}$ nodes} & \rot{Avg. \# of infected $\mathcal{N}$ nodes} \\
    \hline
    \multirow{3}{*}{\textit{$(F \cup T)_{N1}$}}& L & 40 & 30,764 & 33 & 11,340 & 2,005 & 4,302 & 27 & 216\\
    & I & 5 & 246,112 & 267 & 18,909 & 4,486 & 2,997 & 177 & 496\\
    & IP & 287 & 4,288 & 5 & 1,718 & 353 & 982 & 4 & 53\\
    \hline
    \multirow{3}{*}{\textit{$(F \cup T)_{N2}$}}& L & 61 & 47,556 & 82 & 42,135 & 2,893 & 18,601 & 74 & 603\\
    & I & 6 & 483,488 & 836 & 1 & 1 & 1 & 1 & 1   \\
    & IP & 321 & 9,037 & 16 & 2,362 & 284 & 1,759 & 10 & 47\\
    \hline
\multirow{3}{*}{\textit{$(F \cup T)_{N3}$}}& L & 48 & 51,030 & 79 & 29,521 & 3,331 & 10,170 & 72 & 574\\
    & I & 2,647 & 925 & 1 & 952 & 68 & 549 & 1 & 105\\
    & IP & 316 & 7,751 & 12 & 1,488 & 344 & 929 & 7 & 41\\
    \hline
\multirow{3}{*}{\textit{$(F \cup T)_{N4}$}}& L & 41 & 64,961 & 33 & 16,483 & 1,784 & 10,743 & 32 & 230\\
    & I & 1,240 & 2,148 & 1 & 964 & 67 & 645 & 1 & 99 \\
    & IP & 419 & 6,357 & 3 & 2,044 & 242 & 1,387 & 3 & 54\\
    \hline
\multirow{3}{*}{\textit{$(F \cup T)_{N5}$}}& L & 35 & 21,636 & 25 & 14,600 & 2,047 & 4,522 & 23 & 219\\
    & I & 807 & 938 & 1 & 1,075 & 105 & 572 & 1 & 148 \\
    & IP & 142 & 5,333 & 6 & 2,023 & 418 & 1,210 & 5 & 51\\
    \hline
\multirow{3}{*}{\textit{$(F \cup T)_{N6}$}}& L & 31 & 10,574 & 6 & 2,278 &  398 & 1,333 & 5 & 5\\
    & I & 217 & 1,511 & 1 & 545 & 73 & 284 & 1 & 56 \\
    & IP & 115 & 2,850 & 2 & 834 & 150 & 442 & 2 & 31 \\
    \hline
\multirow{3}{*}{\textit{$(F \cup T)_{N7}$}}& L & 27 & 7,188 & 7 & 1,976 &  404 & 664 & 6 & 41\\
    & I & 3 & 64,692 & 61 & 6,504 & 1,590 & 1,535 &  52 & 79\\
    & IP & 78 & 2,488 & 2 & 882 & 217 & 359 & 2 & 27\\
    \hline
\multirow{3}{*}{\textit{$(F \cup T)_{N8}$}}& L & 14 & 114,353 & 15 & 3,801 & 371 & 3,106 & 5 & 28\\
    & I & 3 & 533,649 & 70 & 9,649 & 1,620 & 6,213 &  33 & 95\\
    & IP & 123 & 13,016 & 2 & 755 & 122 & 465 & 2 & 26\\
    \hline
 \multirow{3}{*}{\textit{$(F \cup T)_{N9}$}}& L & 40 & 18,008 & 14 & 4,912 &  964 & 2,089 & 10 & 109\\
    & I & 3 & 240,101 & 192 & 397 & 174 & 194 & 17 & 27\\
    & IP & 192 & 3,752 & 3 & 1,006 & 237 & 479 & 3 & 39\\
    \hline
\multirow{3}{*}{\textit{$(F \cup T)_{N10}$}}& L & 50 & 23,956 & 25 & 6,125 & 898 & 2,915 & 17 & 161\\
    & I & 1,096 & 1,093 & 1 & 576 & 51 & 291 & 1 & 98 \\
    & IP & 228 & 5,253 & 5 & 1,152 & 231 & 709 & 5 & 36\\
    \hline
   \end{tabular}
    }
\end{table*}

\subsection{Evaluation of Metrics}
To measure how good the proposed metrics are able to quantify the vulnerability of nodes and communities, we evaluate the quality of ranking on boundary nodes and communities based on vulnerability scores in comparison with the ground-truth ranking of nodes and communities derived from the news spread in the network. We adopt the ranking evaluation measures widely used in Information Retrieval literature \cite{schutze2008introduction}.

\subsubsection{Evaluate Boundary Node Vulnerability}
A vulnerable boundary node is highly likely to have strong believability with its neighbors. We thus consider the ground truth of a vulnerable node as a node which retweets. The ground truth vulnerability of boundary nodes is binary as we only have information of whether the node retweets or not. 
We thus evaluate this metric using \textit{Average Precision@k} and \textit{Mean Average Precision}. 

\textbf{Average Precision@k (AP@k):} We first compute Precision@k (viz. top-k vulnerable boundary nodes based on the metric as a  percentage of spreader boundary nodes in a community) and then compute the Average Precision@k ($AP@k$) (viz. the average of Precision@k values over all communities in a network).

\textbf{Mean Average Precision (MAP):} Mean Average Precision is computed as the mean of the average precision scores for the top-k boundary nodes over all communities in a network. The formula to compute MAP is given by $\sum_{k=1}^{K} AP(k)/K$, where $K$ denotes total number of communities in the network.

\subsubsection{Evaluate Community Vulnerability}
A community with more number of spreader boundary nodes is more vulnerable to news penetration. As most communities of a network typically have a few spreader boundary nodes, it is not feasible to use node ranking metrics above for evaluating community vulnerability. We thus rank the communities by their vulnerability scores and compare with the ground-truth ranking given by the relative count of spreader boundary nodes in the community.
We use Kendall's tau, which is a correlation measure for ordinal data,  as evaluation metric. 
Kendall's tau close to 1 indicates strong agreement, and that close to -1 indicates strong disagreement between evaluated and ground-truth rankings.

\textbf{Kendall's tau ($\tau$):} Let $rel = [rel_1, rel_2, \dots, rel_n]$ represent the `relevant' ranked list of $n$ communities based on ground-truth vulnerability (quantified as the fraction of boundary nodes that are spreaders), and $ret = [ret_1, ret_2, \dots, ret_n]$ represent the `retrieved' ranked list of communities based on our proposed vulnerability metric. Let $P$ represent the \# of concordant pairs, $Q$ the \# of discordant pairs, $T$ the \# of ties only in $rel$, and $U$ the \# of ties only in $ret$. If a tie occurs for the same pair in both $rel$ and $ret$, it is not added to either $T$ or $U$. Then we calculate $\tau = (P - Q) / sqrt((P + Q + T) * (P + Q + U))$.

\subsection{Results on $DS1$}

\begin{table*}
  \begin{center}
    \caption{Evaluation of vulnerability of boundary nodes for $DS1$.}
    \label{tab:B_DS1}
    \resizebox{\textwidth}{!}{%
    \begin{tabular}{|c|c|c|c|c|c|c|c|c|c|c|c|c|c|c|c|}  \hline
    \multirow{2}{1cm}
    \textbf{} &
    \multicolumn{3}{c|}{\textbf{AP@1}} & 
    \multicolumn{3}{c|}{\textbf{AP@5}} &
    \multicolumn{3}{c|}{\textbf{AP@10}} &
    \multicolumn{3}{c|}{\textbf{AP@15}} &
    \multicolumn{3}{c|}{\textbf{MAP}} \\ \cline{2-16}
    \textbf{} & \textbf{L} & \textbf{I} & \textbf{LP} & 
    \textbf{L} & \textbf{I} & \textbf{LP} & 
    \textbf{L} & \textbf{I} & \textbf{LP} & 
    \textbf{L} & \textbf{I} & \textbf{LP} & 
    \textbf{L} & \textbf{I} & \textbf{LP} \\
      \hline
     M1 & 0.759 & 0.676 & 0.712 & 0.736 & 0.548& 0.519 & 0.606 & 0.543 & 0.533 & 0.661 & 0.505 & 0.566 & 0.672& 0.546 & 0.555  \\
    M2 & 0.818  & 0.749 & 0.907 &  0.769  & 0.733 & 0.799 & 0.821  & 0.699 & 0.999 & 0.733  & 0.666 & 0.999 & 0.785 & 0.733 &  0.875\\
    M3 & 0.805  & 0.642 &0.878  & 0.567  &0.509 & 0.749 & 0.590  & 0.512 & 0.674 & 0.524  &0.586 & 0.833 & 0.596  & 0.577& 0.751 \\
    M4 & 0.468  & 0.714 & 0.750 &  0.366  & 0.674 & 0.633 & 0.323  & 0.523 & 0.659 & 0.325  & 0.454 & 0.799 & 0.350 & 0.569 &0.660 \\ 
    \Xhline{3\arrayrulewidth}
    F1 & 0.892  & 0.749& 0.855 &  0.824  & 0.679& 0.999 & 0.922  &0.499 & 0.799 & 0.899  & 0.422& 0.999 & 0.876  & 0.552& 0.905 \\
    F2 & 0.819 &0.999 & 0.874 &  0.727 & 0.499 & 0.839  & 0.741  & 0.399 &0.924 & 0.706  & 0.266 & 0.999 & 0.714 & 0.518 &  0.900\\ 
    F3 & 0.933 & 0.945 & 0.933 &  0.955 & 0.999 & 0.999 & 0.999 & 0.999 & 0.999 & 0.999 & 0.999 & 0.999 & 0.972 & 0.985 & 0.995 \\
    F4 & 0.999 & 0.999 & 0.999 & 0.955 & 0.999 & 0.999 & {0.979} &  0.999 & 0.999 & {0.999} & 0.999 & 0.999 & {0.991} & 0.999 & 0.999 \\
    \Xhline{3\arrayrulewidth}
    T1 & {0.222} & 0.531 & 0.868 &  {0.424} & 0.492 & 0.716 & {0.439} & 0.349 & 0.479 & {0.377} & 0.344 & 0.533 & {0.450} & 0.424  & 0.644 \\
    T2 & {0.548} & 0.374 & 0.482 &  {0.299} & 0.399 & 0.999 & {0.049} & 0.299 & 0.699 & {0.033} & 0.033 & 0.466 & {0.173} & 0.264 & 0.726 \\
    T3 & {0.666} & 0.470 & 0.913 &  {0.519} & 0.499  & 0.999 & {0.299} & 0.499 & 0.899 & {0.266} & 0.433 & 0.799 & {0.391} & 0.479 & 0.900 \\
    T4 & {0.449} & 0.464 & 0.699 &  {0.399} & 0.000 & 0.479 & {0.409} & 0.000 & 0.499 & {0.362} & 0.000 & 0.366 & {0.399} & 0.106 & 0.500\\
    \hline \hline
        $M_{avg}$ & 0.712  & 0.695 & 0.811 &  0.609  & 0.616 & 0.675 & 0.585 & 0.569 & 0.716 & 0.560  & 0.552 & 0.799 & 0.600 & 0.606 & 0.710\\

$F_{avg}$ & 0.910  & 0.923 & 0.915 & 0.865  & 0.794 & 0.959 &  0.901 & 0.724 & 0.930 & 0.900  & 0.671 & 0.999 & 0.888 & 0.763 & 0.949\\ 

   $T_{avg}$ & 0.471  & 0.459 & 0.740 & 0.410  & 0.347 & 0.798 & 0.299 & 0.286 & 0.644 &  0.259 & 0.202 & 0.541 & 0.353 & 0.318 & 0.692\\
     \hline
    \end{tabular}
    }
  \end{center}
\end{table*}

\begin{table*}
  \begin{center}
        \caption{Evaluation of vulnerability of communities in $DS1$.}
    \label{tab:C_DS1}
        \resizebox{\textwidth}{!}{%
    \begin{tabular}{|l|c|c|c|c|c|c|c|c|c|c|c|c|} 
    \hline
      \textbf{} &  $\tau_{\textbf{M1}}$ & $\tau_{\textbf{M2}}$ & $\tau_{\textbf{M3}}$ & $\tau_{\textbf{M4}}$ & $\tau_{\textbf{F1}}$ & $\tau_{\textbf{F2}}$ & $\tau_{\textbf{F3}}$ & $\tau_{\textbf{F4}}$ & $\tau_{\textbf{T1}}$ & $\tau_{\textbf{T2}}$ & $\tau_{\textbf{T3}}$ & $\tau_{\textbf{T4}}$\\
      \hline
    \textbf{L}  & -0.027 & 0.003  & -0.149 & -0.035  & 0.050 & 0.164 & 0.457 & 0.161 & -0.045 & -0.255 & -0.090 & -0.030\\
    \textbf{I}  & 0.072 & 0.000  & 0.274 & 0.138  & 0.642 & 0.667  & 0.117 & 0.146 & -0.037 & -0.222 & -0.025 & -0.031\\
    \textbf{LP}  & 0.039 & -0.014 & 0.019 & 0.018  & 0.039 & 0.029  & 0.381 & 0.714 & 0.003 & 0.005 & -0.110 & -0.036\\ \hline

    \end{tabular}
    }
  \end{center}
\end{table*}

\begin{figure*}[h]
\centering
\includegraphics[width=\linewidth]{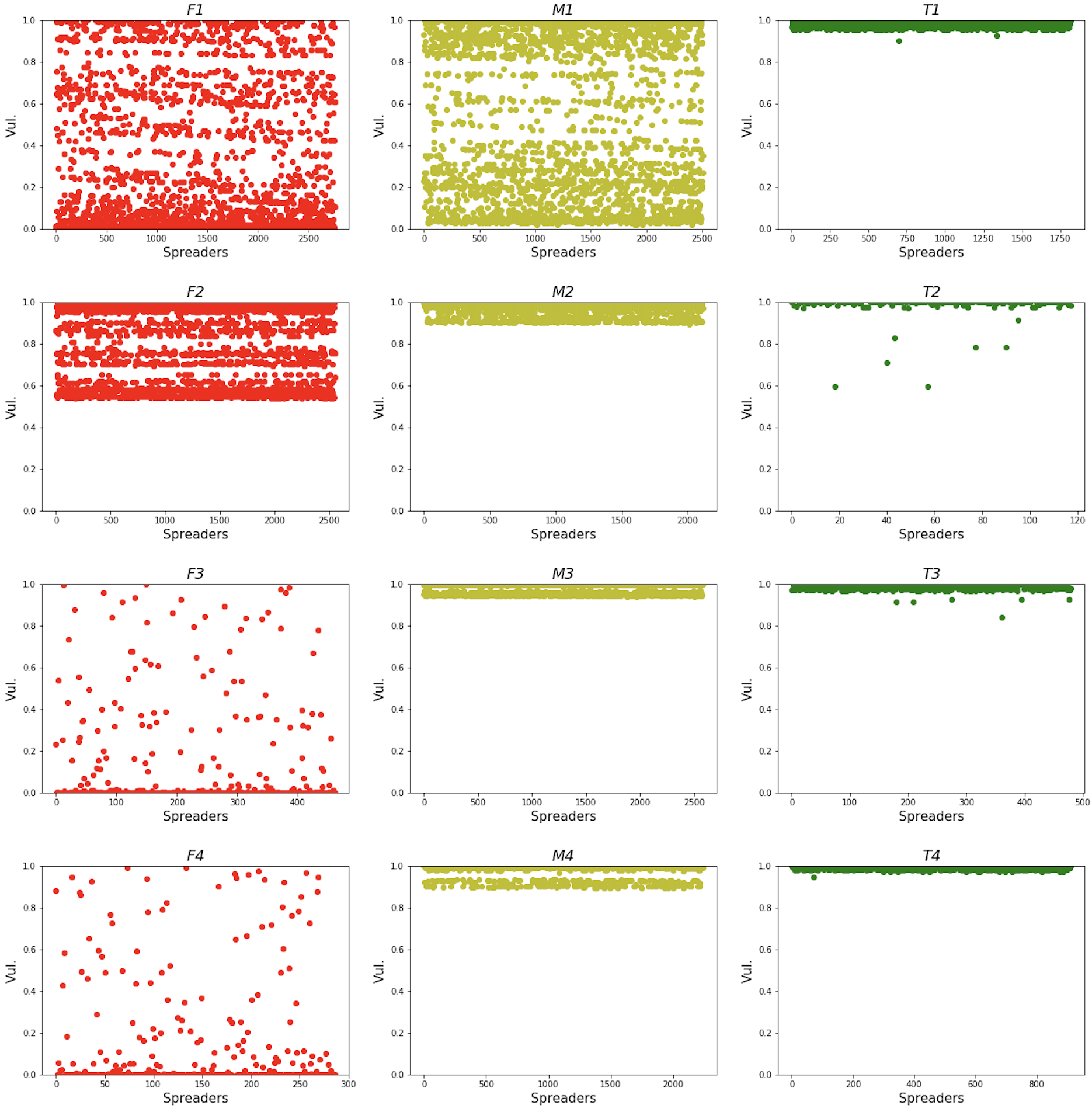}
\caption{Distribution of vulnerability score of spreaders in $DS1$.}
\label{fig:vul_node_DS1}
\end{figure*}

Table~\ref{tab:B_DS1} shows the evaluation results for the proposed metric assessing the vulnerability of boundary nodes for $DS1$. For the twelve networks we show the Average Precision for k = 1, 5, 10 and 15 and compute the MAP for the top-15 results.

AP@1 shows how well we are able to identify the first spreader boundary node based on our metric. Our metric is able to identify the most vulnerable boundary node in AP of 0.712 averaged over the mixture news networks, 0.91 averaged over the false news networks and 0.471 averaged over the true news networks for Louvain; 0.695 averaged over the mixture news networks, 0.923 averaged over the false news networks and 0.459 averaged over the true news  networks for Informap, and 0.811 averaged over the mixture news networks, 0.915 averaged over the false news networks and 0.74 averaged over the true news networks for Label Propagation. Thus, we are able to identify the most vulnerable boundary node of communities in false news networks with average precision of over 90\%. As expected, our metrics show better performance particularly for fake news networks, followed by mixture and then true news networks. Average precision for rest of the k-values also shows similar trend.

Metrics for Louvain-/Infomap-based communities follow a similar trend for the remaining k values. However, Label Propagation communities for k=3 evaluate with AP of 90.25\% averaged over the false news networks, which is over 35\% and 20\% better than the mixture and the true networks, respectively. In this case, true news networks are ranked better than mixture news networks. While k=5 also shows a similar trend, for the rest of the k values Label Propagation-based communities show better performance for the mixture than the true news networks. This insensitivity in evaluation could be attributed to the fact that label propagation algorithm tends to generate more number of communities. Thus, the average community size is much smaller, causing the communities to have sparser boundary and neighbor node sets. 

We also observe that the MAP averaged over the false news networks is 47.86\% better than the mixture and 150\% better than the true news networks for Louvain-based communities; and 25.94\% better than the mixture, and 139.9\% better than the true news networks for Infomap-based  communities; and 33.72\% better than the mixture and 37.14\% better than the true news networks for Label Propagation-based communities. Therefore, we are able to identify most vulnerable boundary nodes of communities in false news networks with an average MAP of over 75\%.

Table~\ref{tab:C_DS1} shows the evaluation results for proposed metric to compute the vulnerability of a community for $DS1$. For the twelve networks the table shows Kendall's tau value $(\tau)$ for communities generated using the three algorithms. We observe that the $\tau$ for mixture and true news networks tend to have a negative correlation with the ground truth community ranking. False news networks on the other hand show a positive correlation, with high values of 0.642, 0.667, 0.457 and 0.714 for $F1$, $F2$, $F3$ and $F4$ respectively. For modular communities generated using Louvain heuristics, our proposed metrics evaluate all false news networks with a positive correlation (average correlation: 0.208) while all true news networks are evaluated with a negative correlation (with average correlation -0.105). Three of the four mixture news networks also have a negative correlation (with average correlation -0.0095). Therefore, our proposed metrics are confirmed to produce better performance on fake news networks, compared to the true and mixture ones.

Figure~\ref{fig:vul_node_DS1} shows the distribution of vulnerability scores of news spreaders of false, mixture and true news networks as per user IDs. We observe that the scores of spreaders in false news networks have more variance (points are more spread out between 0 and 1) than spreaders in mixture news networks. Mixture news networks (except \textit{M1}) 
have less variance, while true news spreaders have least variance. Thus we can conclude that trust-based vulnerability metrics are able to distinguish between spreaders with high and low vulnerability better than true news spreaders (where most spreaders are assigned similar scores). This in turn affects the performance of community vulnerability metrics in a similar way. 

\subsubsection{Case study of \textit{mixture} news spreaders}
\vspace*{-.5cm} 
\begin{figure*}[h]
\centering
\includegraphics[width=\linewidth]{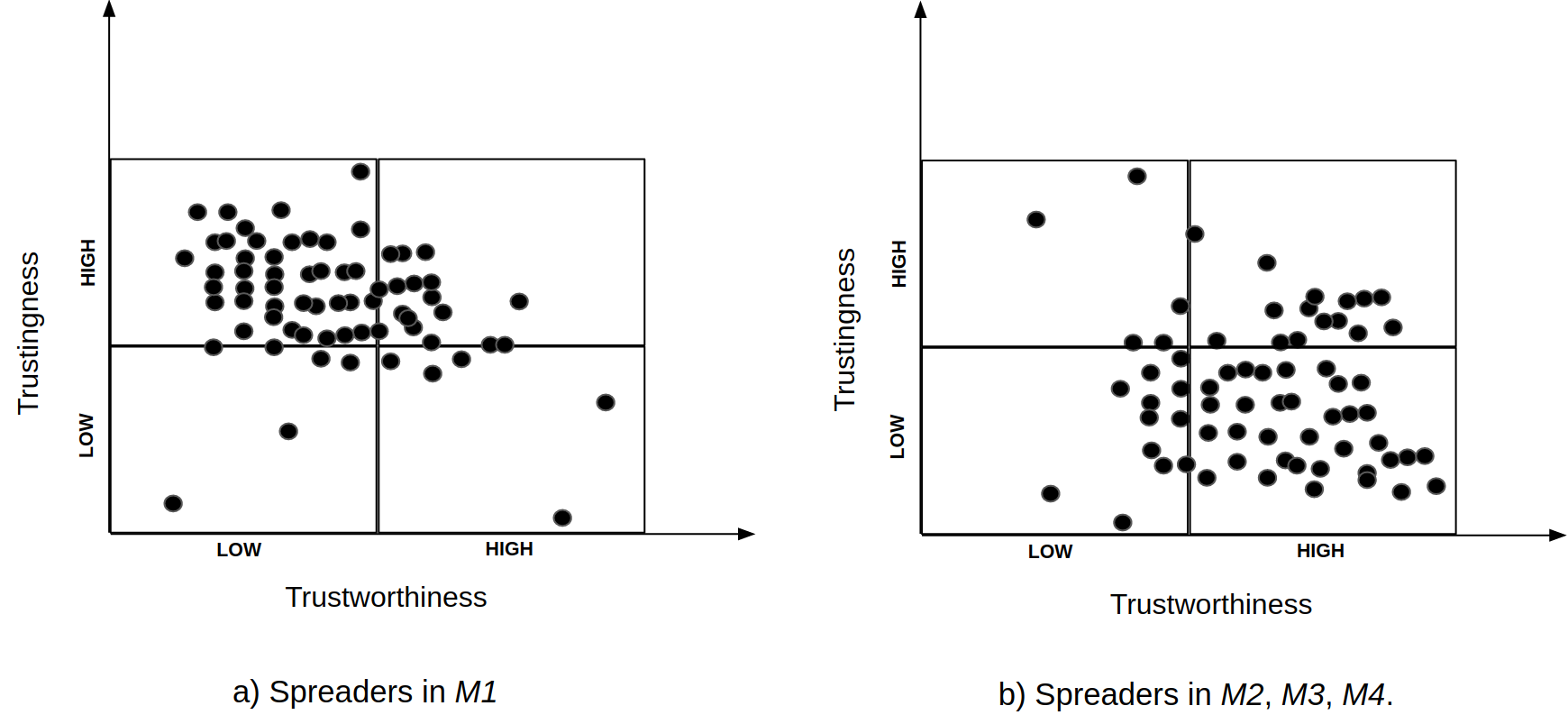}
\caption{Case study of spreaders in $Mixture$ networks.}
\label{fig:case_study}
\end{figure*}
On observing the trustingness and trustworthiness scores of the spreaders of mixture news networks as shown in Figure~\ref{fig:case_study} we notice that most spreaders of \textit{M1} have high trustingness and low trustworthiness scores compared to \textit{M2}, \textit{M3} and \textit{M4} that have low trustingness and high trustworthiness scores. Source of \textit{M1} was tweeted by a conservative with political undertones and it is known that conservatives are more likely to share fake news~\cite{guess2019less}. The information shows  spreading pattern similar to fake news, as  spreaders with high trustingness score shared \textit{M1} without fact checking the claim, unlike the source and spreaders of \textit{M2}, \textit{M3} and \textit{M4} who are not political conservatives.

\subsection{Results on $DS2$}

\begin{table*}
  \begin{center}
    \caption{Evaluation of vulnerability of boundary nodes in $DS2$.}
    \label{tab:B_DS2}
    \resizebox{\textwidth}{!}{%
    \begin{tabular}{|r|c|c|c|c|c|c|c|c|c|c|c|c|c|c|c|}  \hline
    \multirow{2}{1cm}
    \textbf{} &
    \multicolumn{3}{c|}{$\textbf{AP@1}$} & 
    \multicolumn{3}{c|}{$\textbf{AP@5}$} &
    \multicolumn{3}{c|}{$\textbf{AP@10}$} &
    \multicolumn{3}{c|}{$\textbf{AP@15}$} &
    \multicolumn{3}{c|}{\textbf{MAP}} \\ \cline{2-16}
    \textbf{} & \textbf{L} & \textbf{I} & \textbf{LP} & 
    \textbf{L} & \textbf{I} & \textbf{LP} & 
    \textbf{L} & \textbf{I} & \textbf{LP} & 
    \textbf{L} & \textbf{I} & \textbf{LP} & 
    \textbf{L} & \textbf{I} & \textbf{LP} \\
      \hline
     FN1 &0.729&0.999&0.502&0.866&0.533&0.999&0.785&0.333&0.799&0.644&0.222&0.766&0.78&0.48&0.825\\
     TN1 & 0.624&0&0.571&0.819&0&0.999&0.766&0&0.999&0.799&0&0.999&0.766&0&0.872\\
     N1 & 0.799&0.799&0.577&0.799&0.599&0.899&0.789&0.299&0.899&0.752&0.244&0.999&0.789&0.44&0.897\\\hline
     FN2 & 0.728&0.999&0.728&0.599&0&0.933&0.735&0&0.899&0.776&0&0.933&0.711&0.133&0.881\\
     TN2 & 0.702 & 0.314 & 0.62 & 0.616 & 0 & 0.499 & 0.516 & 0 & 0.999 & 0.733 & 0 & 0.999 & 0.565 & 0.079 & 0.831\\
     N2 & 0.745 & 0.999 & 0.669 & 0.614 & 0 & 0.699 & 0.737 & 0 & 0.899 & 0.747 & 0 & 0.933 & 0.697 & 0.133 & 0.806\\\hline
     FN3 & 0.666 & 0.49 & 0.541 & 0.584 & 0.799 & 0.999 & 0.774 & 0.899 & 0.999 & 0.745 & 0.733 & 0.999 & 0.691 & 0.808 & 0.874\\
     TN3 & 0.733 & 0.302 & 0.607 & 0.949 & 0.199 & 0.999 & 0.799 & 0.099 & 0.999 & 0.933 & 0.066 & 0.999 & 0.916 & 0.174 & 0.973\\
     N3 & 0.76 & 0.532 & 0.555 & 0.592 & 0.879 & 0.999 & 0.723 & 0.849 & 0.999 & 0.683 & 0.866 & 0.999 & 0.667 & 0.846 & 0.885\\\hline
     FN4 & 0.599 & 0.999 & 0.556 & 0.699 & 0.299 & 0.999 & 0.585 & 0.099 & 0.899 & 0.866 & 0.066 & 0.999 & 0.637 & 0.282 & 0.878\\
     TN4 & 0.622 & 0.363 & 0.523 & 0.516 & 0 & 0.699 & 0.419 & 0 & 0.599 & 0.666 & 0 & 0.999 & 0.531 & 0.046 & 0.722\\
     N4 & 0.707 & 0.369 & 0.579 & 0.687 & 0.999 & 0.799 & 0.662 & 0.499 & 0.966 & 0.59 & 0.399 & 0.866 & 0.652 & 0.62 & 0.786\\\hline
     FN5 & 0.914 & 0.711 & 0.957 & 0.899 & 0.199 & 0.999 & 0.924 & 0.099 & 0.999 & 0.895 & 0.133 & 0.999 & 0.907 & 0.228 & 0.997\\
     TN5 & 0.599 & 0.999 & 0.824 & 0 & 0.399 & 0.399 & 0 & 0 & 0.199 & 0 & 0 & 0.133 & 0.073 & 0.279 & 0.347\\
     N5 & 0.857 & 0.57 & 0.666 & 0.89 & 0.399 & 0.699 & 0.957 & 0.299 & 0.599 & 0.893 & 0.266 & 0.999 & 0.911 & 0.362 & 0.726\\\hline
     FN6 & 0.769 & 0.499 & 0.762 & 0.519 & 0.533 & 0.999 & 0.499 & 0.599 & 0.899 & 0.466 & 0.533 & 0.666 & 0.549 & 0.581 & 0.867\\
     TN6 & 0.666 & 0.562 & 0.923 & 0 & 0 & 0.599 & 0.099 & 0 & 0 & 0 & 0 & 0 & 0.08 & 0.037 & 0.301\\
     N6 & 0.612 & 0.349 & 0.565 & 0.599 & 0 & 0.699 & 0.533 & 0 & 0.899 & 0.866 & 0 & 0.599 & 0.599 & 0.173 & 0.733\\\hline
     FN7 & 0.749 & 0.499 & 0.914 & 0.399 & 0 & 0.399 & 0.099 & 0 & 0.199 & 0.066 & 0 & 0.133 & 0.285 & 0.033 & 0.314\\
     TN7 & 0.649 & 0.999 & 0.833 & 0.633 & 0.499 & 0.999 & 0.749 & 0.099 & 0.899 & 0.533 & 0.066 & 0.666 & 0.688 & 0.321 & 0.836\\
     N7 & 0.481 & 0.999 & 0.538 & 0.519 & 0.733 & 0.999 & 0.433 & 0.749 & 0.899 & 0.355 & 0.533 & 0.666 & 0.467 & 0.74 & 0.809\\\hline
     FN8 & 0.705 & 0.999 & 0.724 & 0.533 & 0.733 & 0 & 0.499 & 0.566 & 0 & 0.333 & 0.199 & 0 & 0.517 & 0.592 & 0.097\\
     TN8 & 0.499 & 0.499 & 0.721 & 0.499 & 0.499 & 0 & 0 & 0.349 & 0 & 0 & 0 & 0 & 0.218 & 0.352 & 0.048\\
     N8 & 0.499 & 0.999 & 0.442 & 0.733 & 0.933 & 0.199 & 0.499 & 0.599 & 0.099 & 0.333 & 0.422 & 0.066 & 0.546 & 0.713 & 0.2\\\hline
     FN9 & 0.72 & 0.377 & 0.849 & 0.672 & 0 & 0.999 & 0.599 & 0 & 0.999 & 0.483 & 0 & 0.933 & 0.582 & 0.048 & 0.952\\
     TN9 & 0.631 & 0.666 & 0.558 & 0.499 & 0.199 & 0.199 & 0.249 & 0.099 & 0.099 & 0.333 & 0.066 & 0 & 0.409 & 0.215 & 0.133\\
     N9 & 0.724 & 0.333 & 0.526 & 0.619 & 0.199 & 0.899 & 0.539 & 0.099 & 0.699 & 0.516 & 0.066 & 0.999 & 0.568 & 0.154 & 0.817\\\hline
     FN10 & 0.773 & 0.475 & 0.912 & 0.576 & 0.666 & 0.899 & 0.622 & 0.399 & 0.999 & 0.552 & 0.366 & 0.999 & 0.605 & 0.474 & 0.97\\
     TN10 & 0.999 & 0.31 & 0.599 & 0.199 & 0 & 0 & 0.199 & 0 & 0 & 0.133 & 0 & 0 & 0.253 & 0.02 & 0.039\\
     N10 & 0.759 & 0.332 & 0.657 & 0.599 & 0.599 & 0.899 & 0.614 & 0.349 & 0.999 & 0.599 & 0.266 & 0.999 & 0.627 & 0.389 & 0.937\\\hline \hline
     $F_{avg}$ & 0.735 & 0.705 & 0.744 & 0.635 & 0.376 & 0.822 & 0.612 & 0.299 & 0.769 & 0.583 & 0.225 & 0.742 & 0.626 & 0.366 & 0.766\\
     $T_{avg}$ & 0.672 & 0.501 & 0.678 & 0.473 & 0.179 & 0.539 & 0.379 & 0.065 & 0.479 & 0.413 & 0.013 & 0.479 & 0.449 & 0.152 & 0.51\\
     $M_{avg}$ & 0.694 & 0.628 & 0.577 & 0.665 & 0.534 & 0.779 & 0.649 & 0.374 & 0.796 & 0.633 & 0.306 & 0.813 & 0.652 & 0.457 & 0.759\\
     \hline
    \end{tabular}
    }
  \end{center}
\end{table*}

\begin{table*}
  \begin{center}
        \caption{Evaluation of vulnerability of communities in $DS2$.}
    \label{tab:C_DS2}
        \resizebox{\textwidth}{!}{%
    \begin{tabular}{|l|c|c|c|c|c|c|c|c|c|}
     \hline
    \textbf{} & \multicolumn{3}{c|}{$\tau_{F}$} & 
    \multicolumn{3}{c|}{$\tau_{T}$} &
    \multicolumn{3}{c|}{$\tau_{F \cup T}$} \\
    \hline
      \textbf{} &  \textbf{L} & \textbf{I} & \textbf{LP} & \textbf{L} & \textbf{I} & \textbf{LP} &
      \textbf{L} & \textbf{I} & \textbf{LP}\\
      \hline
    \textbf{N1} & 0.009 & 0.333 & 0.075 & 0.171 & 1 & -0.008 & 0.128 & 0 & 0.027\\
    \textbf{N2} & 0.03 & 0.999 & 0.044 & -0.063 & 0.015 & -0.03 & 0.066 & 0.999 & 0.003\\
    \textbf{N3} & -0.351 & -0.001 & 0.012 & 0.2 & -0.04 & 0.06 & -0.411 & -0.012 & 0.044\\
    \textbf{N4} & 0.078 & -1 & -0.009 & -0.222 & 0.065 & -0.011 & -0.051 & -0.007 & -0.022\\
    \textbf{N5} & -0.073 & 0.003 & 0.075 & -0.238 & -1 & 0.01 & -0.055 & 0.02 & 0.051\\
    \textbf{N6} & -0.113 & 1 & 0.039 & -0.055 & 0.017 & 0.052 & -0.092 & 0.033 & -0.038\\
    \textbf{N7} & -0.284 & 0.052 & 0.109 & 0.157 & -1 & -0.062 & -0.065 & 0.333 & 0.066\\
    \textbf{N8} & -0.088 & 0.333 & -0.035 & -0.333 & -1 & -0.089 & -0.076 & -0.333 & 0.011\\
    \textbf{N9} & 0.08 & 0.007 & 0.006 & -0.147 & 0.333 & -0.018 & 0.076 & 0.333 & 0.047\\
    \textbf{N10} & -0.019 & 0.017 & 0.067 & -0.399 & -0.022 & -0.027 & -0.025 & -0.028 & 0.001\\\hline
    \end{tabular}
    }
  \end{center}
\end{table*}


Table~\ref{tab:B_DS2} shows the evaluation results for vulnerability assessment of boundary nodes for $DS1$. For the thirty networks (three each for the ten news events) we show the Average Precision for k = 1, 5, 10 and 15 and compute the MAP for the top-15 results.
Based on the AP@1, we show that our metric is able to identify the most vulnerable boundary node with average precision (aggregated over all news events) of 0.735, 0.672, 0.694 for false, refutation and combined networks repectively when communities are generated using Louvain; 0.705, 0.501, 0.628 when communities are generated using Infomap and 0.744, 0.501, 0.577 when communities are generated using Label propagation method. As in $DS1$, we observe that our propsoed metrics are able to identify spreaders in false information network with higher precision than spreaders in refutation information networks. This can be attributed to the fact that a person's motivation to spread refutation information (whose validity is more certain) is driven more by the nature of the content; unlike false information (whose content is not validated) which is driven less by the content on more by the trust dynamics with the endorser. Metric's performance in identifying false information spreaders in combined network affected slightly due to the presence of refutation information spreading dynamics, but is still better than only refutation information network.

Trends do not drastically vary for other values of k, with Label Propagation performing slightly better than Louvain while Infomap with lowest performance. Also we observe than certain vulnerability scores are drastically low. This can be attributed to the quality of disjoint communities generated by the community detection algorithm. In scenarios where the number of communities is too low or too large, this causes large variation in the boundary and neighbor node count for the community thus affecting the metric score computation.

Through MAP we aggregate the precision scores for top-15 spreader boundary nodes. We observe precison scores of 0.626. 0.366, 0.766 for fale information network; 0.449/ 0.152/ 0.51 for refutation information network; 0.652/ 0.457/ 0.759 for combined network using L/ I/ LP.

Table~\ref{tab:C_DS2} shows the evaluation results for proposed metric to compute the vulnerability of a community for $DS2$. Similar to Table~\ref{tab:C_DS1}, $\tau$ for false information networks tend to have more values greater than zero (i.e. positive correlation) compared to refutation information networks.

\section{Conclusions and Future Work}
We propose novel metrics based on the concept of believability derived from computational trust measures to compute vulnerability of nodes and communities to news spread and show that the metrics is much more sensitive to false information. We confirm our hypothesis that false information have to rely on strong trust among spreaders to propagate while true or refuting information does not. Through experiments on two real-world datasets of large information spreading networks on Twitter we show that our proposed metrics can identify the vulnerable nodes and communities with high precision.
While detection of fake news spreading is a widely studied problem, its containment is not. We believe that the proposed model can be used to identify vulnerable individuals and communities to build content-agnostic fake news spread prevention models. We thus propose the \textit{Community Health Assessment} model as a preliminary idea that exploits the structural characteristics of social networks to identify nodes and communities that are most vulnerable to news spreading. 

As part of future work we would like to extend the proposed ideas to understand the dynamics of news spreading within a community (i.e. through core nodes). We would also like to include temporal features of news spreading into our model.


%
%


\begin{thebibliography}{}
\bibitem{artz2007survey}
Donovan Artz and Yolanda Gil.
\newblock A survey of trust in computer science and the semantic web.
\newblock {\em Web Semantics: Science, Services and Agents on the World Wide
  Web}, 5(2):58--71, 2007.

\bibitem{blondel2008fast}
Vincent~D Blondel, Jean-Loup Guillaume, Renaud Lambiotte, and Etienne Lefebvre.
\newblock Fast unfolding of communities in large networks.
\newblock {\em Journal of statistical mechanics: theory and experiment},
  2008(10):P10008, 2008.

\bibitem{borisyuk2018rosetta}
Fedor Borisyuk, Albert Gordo, and Viswanath Sivakumar.
\newblock Rosetta: Large scale system for text detection and recognition in
  images.
\newblock In {\em Proceedings of the 24th ACM SIGKDD International Conference
  on Knowledge Discovery \& Data Mining}, pages 71--79. ACM, 2018.

\bibitem{quattrociocchi2016echo}
Walter Quattrociocchi, Antonio Scala, and Cass~R Sunstein.
\newblock Echo chambers on facebook.
\newblock {\em Available at SSRN 2795110}, 2016.



\bibitem{fan2013least}
Lidan Fan, Zaixin Lu, Weili Wu, Bhavani Thuraisingham, Huan Ma, and Yuanjun Bi.
\newblock Least cost rumor blocking in social networks.
\newblock In {\em Distributed Computing Systems (ICDCS), 2013 IEEE 33rd
  International Conference on}, pages 540--549. IEEE, 2013.

\bibitem{raghavan2007near}
Usha~Nandini Raghavan, R{\'e}ka Albert, and Soundar Kumara.
\newblock Near linear time algorithm to detect community structures in
  large-scale networks.
\newblock {\em Physical review E}, 76(3):036106, 2007.

\bibitem{rosvall2008maps}
Martin Rosvall and Carl~T Bergstrom.
\newblock Maps of random walks on complex networks reveal community structure.
\newblock {\em Proceedings of the National Academy of Sciences},
  105(4):1118--1123, 2008.

\bibitem{fan2014maximizing}
Lidan Fan, Weili Wu, Xuming Zhai, Kai Xing, Wonjun Lee, and Ding-Zhu Du.
\newblock Maximizing rumor containment in social networks with constrained
  time.
\newblock {\em Social Network Analysis and Mining}, 4(1):214, 2014.

\bibitem{friggeri2014rumor}
Adrien Friggeri, Lada~A Adamic, Dean Eckles, and Justin Cheng.
\newblock Rumor cascades.
\newblock In {\em ICWSM}, 2014.


\bibitem{horne2017just}
Benjamin~D Horne and Sibel Adali.
\newblock This just in: fake news packs a lot in title, uses simpler,
  repetitive content in text body, more similar to satire than real news.
\newblock {\em arXiv preprint arXiv:1703.09398}, 2017.

\bibitem{imran2015processing}
Muhammad Imran, Carlos Castillo, Fernando Diaz, and Sarah Vieweg.
\newblock Processing social media messages in mass emergency: A survey.
\newblock {\em ACM Computing Surveys (CSUR)}, 47(4):67, 2015.

\bibitem{jin2013epidemiological}
Fang Jin, Edward Dougherty, Parang Saraf, Yang Cao, and Naren Ramakrishnan.
\newblock Epidemiological modeling of news and rumors on twitter.
\newblock In {\em Proceedings of the 7th Workshop on Social Network Mining and
  Analysis}, page~8. ACM, 2013.

\bibitem{jin2016news}
Zhiwei Jin, Juan Cao, Yongdong Zhang, and Jiebo Luo.
\newblock News verification by exploiting conflicting social viewpoints in
  microblogs.
\newblock In {\em AAAI}, pages 2972--2978, 2016.

\bibitem{kamvar2003eigentrust}
Sepandar~D Kamvar, Mario~T Schlosser, and Hector Garcia-Molina.
\newblock The eigentrust algorithm for reputation management in p2p networks.
\newblock In {\em Proceedings of the 12th international conference on World
  Wide Web}, pages 640--651. ACM, 2003.

\bibitem{kimura2009efficient}
Masahiro Kimura, Kazumi Saito, and Hiroshi Motoda.
\newblock Efficient estimation of influence functions for sis model on social
  networks.
\newblock In {\em IJCAI}, pages 2046--2051, 2009.


\bibitem{kumar2018false}
Srijan Kumar and Neil Shah.
\newblock False information on web and social media: A survey.
\newblock {\em arXiv preprint arXiv:1804.08559}, 2018.


\bibitem{ma2016detecting}
Jing Ma, Wei Gao, Prasenjit Mitra, Sejeong Kwon, Bernard~J Jansen, Kam-Fai
  Wong, and Meeyoung Cha.
\newblock Detecting rumors from microblogs with recurrent neural networks.
\newblock In {\em IJCAI}, pages 3818--3824, 2016.


\bibitem{ma2017detect}
Jing Ma, Wei Gao, and Kam-Fai Wong.
\newblock Detect rumors in microblog posts using propagation structure via
  kernel learning.
\newblock In {\em Proceedings of the 55th Annual Meeting of the Association for
  Computational Linguistics (Volume 1: Long Papers)}, volume~1, pages 708--717,
  2017.

\bibitem{mishra2011finding}
Abhinav Mishra and Arnab Bhattacharya.
\newblock Finding the bias and prestige of nodes in networks based on trust
  scores.
\newblock In {\em Proceedings of the 20th international conference on World
  wide web}, pages 567--576. ACM, 2011.

\bibitem{mitra2017parsimonious}
Tanushree Mitra, Graham~P Wright, and Eric Gilbert.
\newblock A parsimonious language model of social media credibility across
  disparate events.
\newblock In {\em Proceedings of the 2017 ACM Conference on Computer Supported
  Cooperative Work and Social Computing}, pages 126--145. ACM, 2017.

\bibitem{newman2002spread}
Mark~EJ Newman.
\newblock Spread of epidemic disease on networks.
\newblock {\em Physical review E}, 66(1):016128, 2002.

\bibitem{nguyen2012containment}
Nam~P Nguyen, Guanhua Yan, My~T Thai, and Stephan Eidenbenz.
\newblock Containment of misinformation spread in online social networks.
\newblock In {\em Proceedings of the 4th Annual ACM Web Science Conference},
  pages 213--222. ACM, 2012.

\bibitem{perez2017automatic}
Ver{\'o}nica P{\'e}rez-Rosas, Bennett Kleinberg, Alexandra Lefevre, and Rada
  Mihalcea.
\newblock Automatic detection of fake news.
\newblock {\em arXiv preprint arXiv:1708.07104}, 2017.


\bibitem{potthast2017stylometric}
Martin Potthast, Johannes Kiesel, Kevin Reinartz, Janek Bevendorff, and Benno
  Stein.
\newblock A stylometric inquiry into hyperpartisan and fake news.
\newblock {\em arXiv preprint arXiv:1702.05638}, 2017.


\bibitem{rath2017retweet}
Bhavtosh Rath, Wei Gao, Jing Ma, and Jaideep Srivastava.
\newblock From retweet to believability: Utilizing trust to identify rumor
  spreaders on twitter.
\newblock In {\em Proceedings of the 2017 IEEE/ACM International Conference on
  Advances in Social Networks Analysis and Mining 2017}, pages 179--186. ACM,
  2017.

\bibitem{roy2015computational}
Atanu Roy.
\newblock Computational trust at various granularities in social networks.
\newblock 2015.

\bibitem{roy2016trustingness}
Atanu Roy, Chandrima Sarkar, Jaideep Srivastava, and Jisu Huh.
\newblock Trustingness \& trustworthiness: A pair of complementary trust
  measures in a social network.
\newblock In {\em Advances in Social Networks Analysis and Mining (ASONAM),
  2016 IEEE/ACM International Conference on}, pages 549--554. IEEE, 2016.


\bibitem{schutze2008introduction}
Hinrich Sch{\"u}tze, Christopher~D Manning, and Prabhakar Raghavan.
\newblock {\em Introduction to information retrieval}, volume~39.
\newblock Cambridge University Press, 2008.

\bibitem{shah2011rumors}
Devavrat Shah and Tauhid Zaman.
\newblock Rumors in a network: Who's the culprit?
\newblock {\em IEEE Transactions on information theory}, 57(8):5163--5181,
  2011.

\bibitem{sherchan2013survey}
Wanita Sherchan, Surya Nepal, and Cecile Paris.
\newblock A survey of trust in social networks.
\newblock {\em ACM Computing Surveys (CSUR)}, 45(4):47, 2013.


\bibitem{vosoughi2018spread}
Soroush Vosoughi, Deb Roy, and Sinan Aral.
\newblock The spread of true and false news online.
\newblock {\em Science}, 359(6380):1146--1151, 2018.


\bibitem{yang2012automatic}
Fan Yang, Yang Liu, Xiaohui Yu, and Min Yang.
\newblock Automatic detection of rumor on sina weibo.
\newblock In {\em Proceedings of the ACM SIGKDD Workshop on Mining Data
  Semantics}, page~13. ACM, 2012.

\bibitem{zhao2013sir}
Laijun Zhao, Hongxin Cui, Xiaoyan Qiu, Xiaoli Wang, and Jiajia Wang.
\newblock Sir rumor spreading model in the new media age.
\newblock {\em Physica A: Statistical Mechanics and its Applications},
  392(4):995--1003, 2013.

\bibitem{zhao2009and}
Dejin Zhao and Mary~Beth Rosson.
\newblock How and why people twitter: the role that micro-blogging plays in
  informal communication at work.
\newblock In {\em Proceedings of the ACM 2009 international conference on
  Supporting group work}, pages 243--252, 2009.
  
\bibitem{zhu2016information}
Kai Zhu and Lei Ying.
\newblock Information source detection in the sir model: A sample-path-based
  approach.
\newblock {\em IEEE/ACM Transactions on Networking (TON)}, 24(1):408--421,
  2016.

\bibitem{ziegler2004spreading}
C-N Ziegler and Georg Lausen.
\newblock Spreading activation models for trust propagation.
\newblock In {\em e-Technology, e-Commerce and e-Service, 2004. EEE'04. 2004
  IEEE International Conference on}, pages 83--97. IEEE, 2004.
  
  \bibitem{sharma2019combating}
Sharma, Karishma and Qian, Feng and Jiang, He and Ruchansky, Natali and Zhang, Ming and Liu, Yan
\newblock Combating fake news: A survey on identification and mitigation
  techniques.
\newblock {\em ACM Transactions on Intelligent Systems and Technology (TIST)}, 2019.

\bibitem{khattar2019mvae}
Dhruv Khattar, Jaipal~Singh Goud, Manish Gupta, and Vasudeva Varma.
\newblock Mvae: Multimodal variational autoencoder for fake news detection.
\newblock In {\em The World Wide Web Conference}, pages 2915--2921, 2019.

\bibitem{lu2020gcan}
Yi-Ju Lu and Cheng-Te Li.
\newblock Gcan: Graph-aware co-attention networks for explainable fake news
  detection on social media.
\newblock {\em arXiv preprint arXiv:2004.11648}, 2020.

\bibitem{ma2019detect}
Jing Ma, Wei Gao, and Kam-Fai Wong.
\newblock Detect rumors on twitter by promoting information campaigns with
  generative adversarial learning.
\newblock In {\em The World Wide Web Conference}, pages 3049--3055, 2019.

\bibitem{shu2019defend}
Kai Shu, Limeng Cui, Suhang Wang, Dongwon Lee, and Huan Liu.
\newblock defend: Explainable fake news detection.
\newblock In {\em Proceedings of the 25th ACM SIGKDD International Conference
  on Knowledge Discovery \& Data Mining}, pages 395--405, 2019.

\bibitem{zhang2020fakedetector}
Jiawei Zhang, Bowen Dong, and S~Yu Philip.
\newblock Fakedetector: Effective fake news detection with deep diffusive
  neural network.
\newblock In {\em 2020 IEEE 36th International Conference on Data Engineering
  (ICDE)}, pages 1826--1829. IEEE, 2020.
  
\bibitem{chen2020modeling}
Tinggui Chen, Jiawen Shi, Jianjun Yang, Guodong Cong, and Gongfa Li.
\newblock Modeling public opinion polarization in group behavior by integrating
  sirs-based information diffusion process.
\newblock {\em Complexity}, 2020, 2020.

\bibitem{jin2013epidemiological}
Fang Jin, Edward Dougherty, Parang Saraf, Yang Cao, and Naren Ramakrishnan.
\newblock Epidemiological modeling of news and rumors on twitter.
\newblock In {\em Proceedings of the 7th workshop on social network mining and
  analysis}, pages 1--9, 2013.

\bibitem{khelil2002epidemic}
Abdelmajid Khelil, Christian Becker, Jing Tian, and Kurt Rothermel.
\newblock An epidemic model for information diffusion in manets.
\newblock In {\em Proceedings of the 5th ACM international workshop on Modeling
  analysis and simulation of wireless and mobile systems}, pages 54--60, 2002.

\bibitem{liu2016characterizing}
Yu~Liu, Bai Wang, Bin Wu, Suiming Shang, Yunlei Zhang, and Chuan Shi.
\newblock Characterizing super-spreading in microblog: An epidemic-based
  information propagation model.
\newblock {\em Physica A: Statistical Mechanics and its Applications},
  463:202--218, 2016.

\bibitem{rui2018spir}
Xiaobin Rui, Fanrong Meng, Zhixiao Wang, Guan Yuan, and Changjiang Du.
\newblock Spir: The potential spreaders involved sir model for information
  diffusion in social networks.
\newblock {\em Physica A: Statistical Mechanics and its Applications},
  506:254--269, 2018.

\bibitem{hoangvirality}
Tuan-Anh HOANG and Ee~Peng LIM.
\newblock Virality and susceptibility in information diffusions.
\newblock In {\em Proceedings of the Sixth International Conference on Weblogs
  and Social Media, 2012}.

\bibitem{guess2019less}
Andrew Guess, Jonathan Nagler, and Joshua Tucker.
\newblock Less than you think: Prevalence and predictors of fake news
  dissemination on facebook.
\newblock {\em Science advances}, 5(1):eaau4586, 2019.

\bibitem{shen2019gullible}
Tracy~Jia Shen, Robert Cowell, Aditi Gupta, Thai Le, Amulya Yadav, and Dongwon
  Lee.
\newblock How gullible are you? predicting susceptibility to fake news.
\newblock In {\em Proceedings of the 10th ACM Conference on Web Science}, pages
  287--288, 2019.
  

\end{thebibliography}


\end{document}